\documentclass[journal,final,twocolumn,10pt,twoside]{IEEEtran}

\IEEEoverridecommandlockouts
\usepackage[noadjust]{cite}
\usepackage{amsmath,amssymb,amsfonts,amsbsy,amsthm}
\usepackage{algorithmic}    
\usepackage{graphicx}
\usepackage{textcomp}
\usepackage{xcolor}
\usepackage{siunitx}
\usepackage{balance}
\usepackage[caption=false,font=footnotesize]{subfig} 

\def\BibTeX{{\rm B\kern-.05em{\sc i\kern-.025em b}\kern-.08em
    T\kern-.1667em\lower.7ex\hbox{E}\kern-.125emX}}

\definecolor{mblue}{rgb}{0,0.1,1.0}
\definecolor{mred}{rgb}{0.9,0.1,0.1}

\DeclareMathOperator*{\argmax}{arg\,max}
\DeclareMathOperator*{\argmin}{arg\,min}

\begin{document}

\title{\huge Higher Order Derivative-Based Receiver Pre-processing for Molecular Communications \\
		\thanks{This paper was presented in part at the IEEE Global Communications Conference (GLOBECOM 2020) \cite{gursoymitra_globecom}, and in part at the ACM International Conference on Nanoscale Computing and Communication (NanoCom) \cite{gursoymitra_ACM}. This work has been funded in part by one or more of the following grants: ONR N00014-15-1-2550, NSF CCF-1817200, ARO W911NF1910269, Cisco Foundation 1980393, DOE DE-SC0021417, Swedish Research Council 2018-04359, NSF CCF-2008927, ONR 503400-78050.}
}

%\and
%\and
%\and
%\and
%email address or ORCID}*
%\and

\author{\large{Mustafa Can Gursoy and Urbashi Mitra}
	\\ \normalsize Department of Electrical and Computer Engineering, University of Southern California, Los Angeles, CA, USA
	\\ E-mail: \{mgursoy, ubli\}@usc.edu
}

\markboth{Submitted to IEEE Transactions on Molecular, Biological, and Multi-Scale Communications}
{Submitted to IEEE Transactions on Molecular, Biological, and Multi-Scale Communications}

\maketitle

\begin{abstract}

While molecular communication via diffusion experiences significant inter-symbol interference (ISI), recent work suggests that ISI can be mitigated via time differentiation pre-processing which achieves pulse narrowing. Herein, the approach is generalized to higher order differentiation. The fundamental trade-off between ISI mitigation and noise amplification is characterized, showing the existence of an optimal derivative order that minimizes the bit error rate (BER). Theoretical analyses of the BER and a signal-to-interference-plus-noise ratio are provided, the derivative order optimization problem is posed and solved for threshold-based detectors. For more complex detectors which exploit a window memory, it is shown that derivative pre-processing can strongly reduce the size of the needed window. Extensive numerical results confirm the accuracy of theoretical derivations, the gains in performance via derivative pre-processing over other methods and the impact of the optimal derivative order.  Derivative pre-processing offers a low complexity/high-performance method for reducing ISI at the expense of increased transmission power to reduce noise amplification.

%Molecular communication via diffusion (MCD) faces high levels of inter-symbol interference (ISI), which lowers the data rate at which reliable MCD links can be realized. 
%Inspired by the recent studies on achieving ISI mitigation through time differentiation, this paper proposes to use higher order derivatives to pre-process the signal at the receiver. For a derivative-based receiver, the fundamental trade-off between ISI mitigation and noise amplification is characterized, which implies the existence of an optimal derivative order that minimizes the bit error ratio (BER). Through theoretical BER expressions and a signal-to-interference and noise ratio-like (SINR) objective function, the derivative order optimization problem is addressed for threshold-based detectors. For more complex detectors that use memory, it is shown that the strong ISI mitigation of the derivative operator greatly reduces the memory window length requirements for reliable operation. Obtained numerical results confirm the accuracy of the derived BER expressions, the efficacy of SINR in optimizing the derivative order, and the error performance improvement introduced by higher order derivatives. Overall, the derivative-based pre-processor provides a computationally cheap strategy that achieves higher data rates with low error rates, at a cost of increased transmission power to combat the noise amplification.
\end{abstract}

%\vspace{-0.5cm}

\section{Introduction}
\par Molecular communication via diffusion (MCD) enables communication through the emission of chemical (molecular) signals \cite{molecular_1}. In an MCD system, the information is encoded into a physical property of the molecular signal such as its emission intensity \cite{CSKMOSK}, emitted molecule type \cite{isomermosk}, time of emission \cite{molcom_PPM}, spatial location of emission \cite{MSSK_TCOM,spatial_mod_huang}, or a combination of these signaling degrees of freedom \cite{MSSK_TCOM,spatial_mod_huang,MCSK,MCPM,typebasedsign_MNK}. 
After their emission from the transmitter, the messenger molecules randomly propagate in the fluid communication medium, exhibiting Brownian motion \cite{book_eckford}. This stochasticity causes some molecules to never arrive at the receiver, and creates delays in some molecules that do arrive. From a communications engineering perspective, the molecules that arrive later than intended cause inter-symbol interference (ISI). ISI is the leading cause of the notoriously low data rates of MCD.

\par The ISI problem has been tackled through both transmitter and receiver side solutions. In particular, single- or multi-molecule modulation schemes have been considered \cite{mod_survey}, source and channel codes have been designed \cite{MCG_MNK_UM_survey}, as well as transmitter-side pre-equalization approaches \cite{burcu_pre,typebasedsign_MNK}. At the receiver side, the maximum {\em a posteriori} (MAP) and maximum likelihood (ML) sequence detectors are considered by \cite{receiver_design}, as well as decision feedback and minimum mean squared error (MMSE) equalizers that account for ISI. Inspired by the computational constraints of a nano-machine, a low complexity, adaptive threshold detector is presented in \cite{ATD_2016}. In \cite{MLDA_mitra_2014}, a decision feedback mechanism is utilized to estimate ISI and aid a symbol-by-symbol detector (memory limited decision aided decoder, MLDA). A similar decision feedback mechanism is also used in \cite{sprt_mitra} in the context of a sequential probability ratio test-based MCD detector. 

\par Recently, it was shown in \cite{lin_derivative} that applying a single discrete-time derivative on the received signal mitigates ISI in concentration-based synthetic MCD. Subsequently, a rising-edge-based detection with differentiation strategy was devised in \cite{yu_risingedge} for macro-scale molecular MIMO.  In \cite{lin_derivative}, it was shown that a single differentiation narrowed the received signal pulse thus mitigating ISI.  Herein, we consider {\bf multiple} orders of differentiation and their pairing with a variety of detector strategies as noted below. It should be observed that differentiation is not only an engineered mechanism for micro-or nano-machines, but is a processing/sensing that occurs in organisms. In particular, bacterial responses are affected by the rate of change of physical and biochemical quantities. Examples include detecting spatial gradients of bio-molecule concentration used for \textit{chemotaxis} \cite{chemotaxis} and the varying effect of heating rate in protein synthesis \cite{ecoli_heatrate,ecoli_heatrate2}.

%\par \tcan{Whether through spatial gradients of bio-molecule concentration used for chemotaxis \cite{chemotaxis}, or the effects of heating rate in protein synthesis \cite{ecoli_heatrate,ecoli_heatrate2}, bacterial responses affected by the rate of change of physical and biochemical quantities constantly occur in the microbial world. From these prior works, it can be inferred that compared to the original signal, the derivative signal has a smaller peak time and a more rapid decay after the peak that induces less ISI for consequent transmissions.

%\par In this paper, we extend these prior works, and generalize them to an arbitrary derivative order $m$. 

\par In our preliminary study \cite{gursoymitra_globecom}, we had introduced the higher order derivative concept, discussed its fundamental trade-off between ISI mitigation and noise amplification, and introduced a lower complexity, banded alternative to the optimal maximum likelihood sequence detector that exploits the ISI mitigation offered by higher order differentiation. In addition, in a separate preliminary study \cite{gursoymitra_ACM}, we had improved the fixed threshold detector used by \cite{lin_derivative} for $m=1$ and \cite{gursoymitra_globecom} for $m\geq1$, and provided an objective function to optimize the derivative order $m$ using the new fixed threshold detector. This paper extends and completes these two works by providing complete derivations and proofs, deriving the theoretical bit error ratio (BER) of the detector proposed in \cite{gursoymitra_ACM}, introducing a new detector to be paired with the derivative operator, as well as addressing the computational complexities of the detectors and the asymptotic relationship between derivative orders. The contributions of this paper are as follows:
\begin{enumerate}
    \item We generalize the initial endeavors of \cite{lin_derivative} to a pre-processor with an arbitrary derivative order $m$.
    %and provide the derivative-based pre-processing framework.
    \item We characterize the fundamental trade-off of the derivative-based pre-processing framework between ISI mitigation and noise amplification. %This trade-off implies the existence of an optimal derivative order that minimizes error.
    \item Framing the derivative operation as a receiver pre-processor block that takes place before detection, we present several derivative operator-detector pairs. To this end, we provide the limited-memory, banded version of the MLSD, generalize the MLDA to an arbitrary derivative order, and adapt two threshold-based detectors to the derivative pre-processor. 
    \item We derive the theoretical bit error ratio (BER) expressions for the threshold-based detectors.
    \item We provide a signal-to-interference-plus-noise ratio-like (SINR) objective function that is compatible with an arbitrary derivative order. Through this objective function and the theoretical error expression, we address the derivative order optimization problem.
    \item Obtained numerical results demonstrate the characterized trade-off between ISI mitigation and noise amplification, and with proper derivative order optimization, confirm the performance improvement of the derivative operator. 
    %The results suggest that given the transmitter can handle large transmission powers, the derivative pre-processor can achieve fast and reliable MCD, while still keeping the receiver complexity low.
\end{enumerate}

\par The rest of the paper is organized as follows: Section \ref{sec:channel} presents the MCD channel model under consideration. Section \ref{sec:proposed} proposes the $m^{th}$ order derivative operator and discusses the fundamental trade-off between ISI mitigation and noise amplification. Section \ref{sec:detector} introduces possible detectors to be combined with the derivative-based pre-processor, discussing their main strategies of operation and computation complexities. Section \ref{sec:costfunction} addresses the derivative order optimization problem through theoretical BER expressions and an alternative objective function. Section \ref{sec:results} presents the comparative numerical results, and Section \ref{sec:conclusion} concludes the paper.

%\vspace{-0.2cm}
\section{System Model}
\label{sec:channel}

\par
\begin{figure}[!t]
	\centering
	\includegraphics[width=0.37\textwidth]{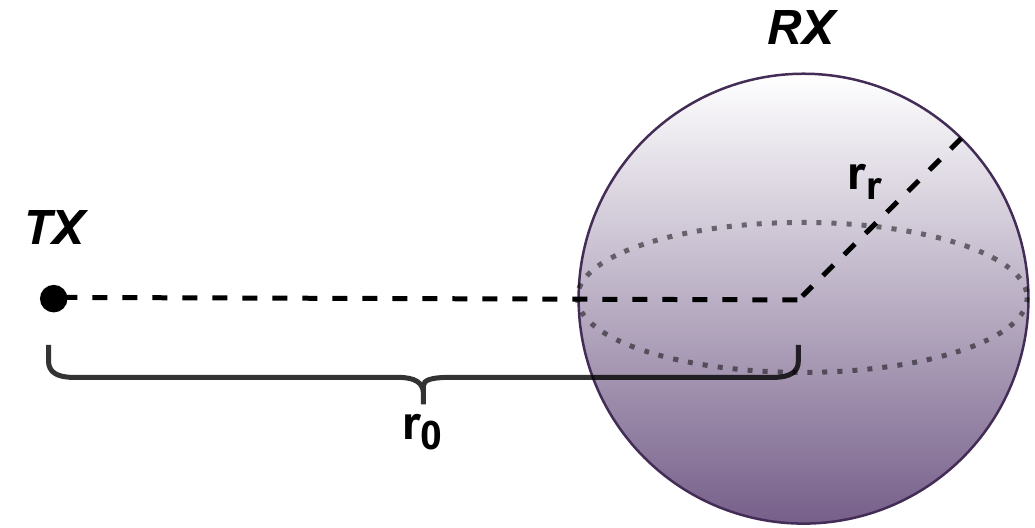} 
	\caption{The considered system model of interest.}
	\label{fig:topology}
	%\vspace*{-0.1in}
\end{figure}

 In this paper, the considered topology consists of a point transmitter and a spherical absorbing receiver in a $3$-D, unbounded environment. The distance between the transmitter and the center of the spherical receiver is denoted by $r_0$ and the radius of the receiver is denoted by $r_r$. Overall, the considered topology is presented in Figure \ref{fig:topology}.

\par For the topology presented in Figure \ref{fig:topology}, denoting the diffusion coefficient of the messenger molecules by $D$, the time density of molecule arrivals (\textit{i.e.,} the channel impulse response, CIR) is presented in \cite{3Dchar} to be
\begin{equation} \label{eq:arrival_pdf}
	f_{\textrm{hit}}(t) = \frac{r_r}{r_0} \frac{1}{\sqrt[]{4\pi Dt}} \frac{r_0-r_r}{t} e^{- \frac{(r_0-r_r)^2}{4Dt} }, \hspace{0.35cm} t \in (0,\infty),
\end{equation}
with its time integral being equal to
\begin{equation}\label{eq:arrival_cdf}
F_{\textrm{hit}}(t) = \frac{r_{r}}{r_{0}} \text{erfc}\bigg( \frac{r_{0}-r_{r}}{\sqrt[]{4Dt}} \bigg).
\end{equation}

\par Note that Equation \eqref{eq:arrival_cdf} represents the probability of a molecule's arrival at the receiver up to time $t$. In this paper, we consider a time-slotted MCD system where the transmitter and receiver are perfectly synchronized. Using \eqref{eq:arrival_cdf}, the entries of the channel coefficient vector $\boldsymbol{h}$ can be obtained by 
\begin{equation}
    \label{eq:h_vector}
	h[n] = F_{\textrm{hit}}(n t_s) - F_{\textrm{hit}}\left((n-1) t_s\right) , \hspace{0.5cm} n = 1, 2, \dots, LN,
\end{equation}
where $t_s$ is the duration of a time slot (sample), $N$ is the number of samples per one symbol duration (\textit{i.e.,} $t_{\textrm{symbol}} = N t_s$), and $L$ denotes the length of the channel memory window in symbols. 

\par Throughout the paper, we consider binary concentration shift keying (BCSK, \cite{CSKMOSK}) signaling with equiprobable symbol transmissions, which defines transmitting a bit-$1$ by emitting $M$ molecules, and a bit-$0$ by emitting no molecules. Note that since BCSK is a binary modulation scheme, the symbol duration is equal to the bit duration ($t_{\textrm{symbol}} = t_b$). Herein, we denote $\boldsymbol{s}$ as the binary vector of transmitted bits. Employing BCSK, assuming an idealized transmitter and that the emission occurs at the beginning of the symbol interval, the emission count vector $\boldsymbol{x}$ is given by
%Furthermore, we assume an idealized transmitter. That is, denoting $\boldsymbol{s}$ as the binary vector of transmitted bits and $\boldsymbol{x}$ as the vector that holds the emitted number of molecules, $\boldsymbol{x}$ is deterministic with respect to $\boldsymbol{s}$. In particular to BCSK, considering the emission at the beginning of the symbol interval, $\boldsymbol{x}$ is characterized as 
\begin{equation}
    \label{eq:bcsk}
    x[i] = 
    \begin{cases}
      M, & \text{if $s[k] = 1$ and $i = (k-1)N + 1$} \\
      0, & \text{otherwise}.
    \end{cases}
\end{equation}

\par Given the channel coefficient vector $\boldsymbol{h}$ and the emission count vector $\boldsymbol{x}$, the $n^{th}$ sample of the received signal $\boldsymbol{y}$ can be approximated as a Poisson distributed random variable \cite{arrivalmodel}:
\begin{equation} \label{eq:arrival_poisson}
	y[n] \sim \mathcal{P} \left( \lambda_s + \sum_{k = 1}^{LN} h[k] x[n-k+1] \right),
\end{equation} 
where $\lambda_s$ is the rate of the external Poisson noise. This model is also referred to as the linear time-invariant (LTI)-Poisson channel \cite{LTI_Poisson}. Herein, we employ the Gaussian approximation of the Poisson arrival counts \cite{arrivalmodel}. Therefore, for transmissions in blocks of length $S$, and separating the deterministic and random components of $\boldsymbol{y} \sim (\boldsymbol{\mu},\boldsymbol{\Sigma})$, the received signal, in vector form, can be expressed as
\begin{equation} \label{eq:matrix_form}
	\begin{split}
	\boldsymbol{y} = (\boldsymbol{H}\boldsymbol{x} + \lambda_s \boldsymbol{j}) + \boldsymbol{\eta}.
	\end{split}
\end{equation}
Here, $\boldsymbol{H}$ denotes the $SN \times SN$ Toeplitz matrix corresponding to the convolution operation of LTI-Poisson in \eqref{eq:arrival_poisson}, $\boldsymbol{j}$ is an $SN \times 1$ vector of ones, and $\boldsymbol{\eta} \sim \mathcal{N} \left( \boldsymbol{0}, \boldsymbol{\Sigma} \right)$, where
\begin{equation} \label{eq:matrix_form2}
    \boldsymbol{\Sigma} = \textrm{diag}\{\boldsymbol{H}\boldsymbol{x}\} + \lambda_s \boldsymbol{I}.
\end{equation}
Note that $\boldsymbol{\Sigma}$ is dependent on $\boldsymbol{s}$ through $\boldsymbol{x}$, which implies the signal-dependent noise phenomenon of MCD systems \cite{receiver_design,typebasedsign_MNK,jamali_mf,MCG_MNK_UM_survey}.

%%%%%%%%%%%%%%%%%%%%%%

\begin{figure*}[!t]
	\centering
	\includegraphics[width=0.98\textwidth]{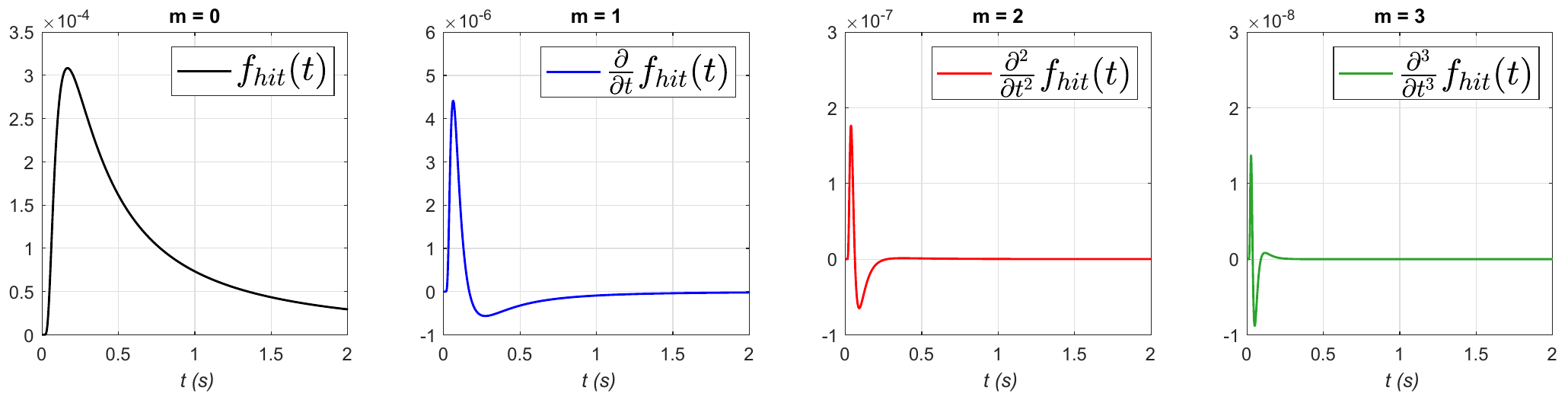} 
	\caption{Evolution of $\frac{\partial^{m} f_{hit}}{\partial t^{m}}$ with the derivative order $m$. $r_0 = \SI{15}{\micro\meter}$, $r_r = \SI{5}{\micro\meter}$, $D = 100 \frac{\SI{}{\micro\meter\squared}}{\SI{}{\second}}$.}
	\label{fig:fhits} 
\end{figure*}

\section{Fundamentals of Derivative-Based Pre-Processing}
\label{sec:proposed}

\par Herein, we discuss the main motivation and key statistical properties of the $m^{th}$ order derivative operator in an MCD system. To this end, we first address how the CIR in \eqref{eq:arrival_pdf} evolves with the derivative order $m$. Recalling a result from our prior work, the first peak time of $\frac{\partial^{m} f_{\textrm{hit}}}{\partial t^{m}}$ (\textit{i.e.,} the time at which the $m^{th}$ derivative of the CIR achieves its first maximum) is a monotonically decreasing function of the derivative order $m$, \cite[Proposition 1]{gursoymitra_globecom}. Furthermore, $\frac{\partial^{m} f_{\textrm{hit}}}{\partial t^{m}}$ shrinks in pulse width with increasing $m$, as clearly observed in Figure~\ref{fig:fhits}.

\par From a receiver design standpoint, the consequence of the above two phenomena is an effective narrowing of each emitted pulse at the receiver, which mitigates ISI for consecutive symbol transmission scenarios. In order to characterize this effect for the time-slotted, discrete time channel, we define the discrete-time forward derivative operator, denoted by $\boldsymbol{D}$, as
\begin{equation} \label{eq:Dmatrix}
	\boldsymbol{D} = \begin{bmatrix}
		-1 & 1 & 0  & \cdots & 0 \\
		0 & -1 & 1  & \cdots & 0 \\
		\vdots & \vdots & \ddots & \ddots & \vdots \\
		\vdots & \vdots &  & -1 & 1 \\
		0 & 0 & \cdots  & 0 &-1
	\end{bmatrix}.
\end{equation}
Furthermore, we denote the output of the $m^{th}$ order derivative operator as $\boldsymbol{y}_{(m)} \sim \mathcal{N}(\boldsymbol{\mu}_{(m)},\boldsymbol{\Sigma}_{(m)})$. Then, $\boldsymbol{y}_{(m)}$ can be expressed as
\begin{equation} 
    \label{eq:y_derivatived}
	\begin{split}
		\boldsymbol{y}_{(m)} &= \boldsymbol{D}^m \boldsymbol{y}  \\
		&= \boldsymbol{D}^m \left(\boldsymbol{H}\boldsymbol{x} + \lambda_s \boldsymbol{j} \right) + \boldsymbol{D}^m  \boldsymbol{\eta}\\
		&\sim \mathcal{N}\left( \boldsymbol{D}^m \boldsymbol{\mu}, \boldsymbol{D}^m \boldsymbol{\Sigma} (\boldsymbol{D}^{\top})^m \right) .
	\end{split}
\end{equation}
The mean of $\boldsymbol{y}_{(m)}$ reflects the aforementioned ISI mitigation introduced by the $\boldsymbol{D}^m$ operator. However, since $\boldsymbol{\Sigma}_{(m)} = \boldsymbol{D}^m \boldsymbol{\Sigma} (\boldsymbol{D}^{\top})^m$, the application of the $m^{th}$ order derivative operator inherently introduces noise amplification and coloration into the received signal. We note that the banded diagonal form of $\boldsymbol{D}^m$ increases in both width and magnitude with increasing $m$, further increasing the coloration and amplification. 

\par Overall, increasing $m$ results in better ISI mitigation at the cost of a more severe enhancement of the noise power. 
%As also discussed in \cite{gursoymitra_globecom,gursoymitra_ACM}, 
This interplay between ISI mitigation and noise amplification implies a fundamental trade-off for a derivative-based MCD receiver, implying the existence of an optimal derivative order $m$ that minimizes the error probability. We will address this optimization problem in Section \ref{sec:costfunction}.

\section{Detector Design}
\label{sec:detector}

\par
\begin{figure*}[!t]
	\centering
	\includegraphics[width=0.8\textwidth]{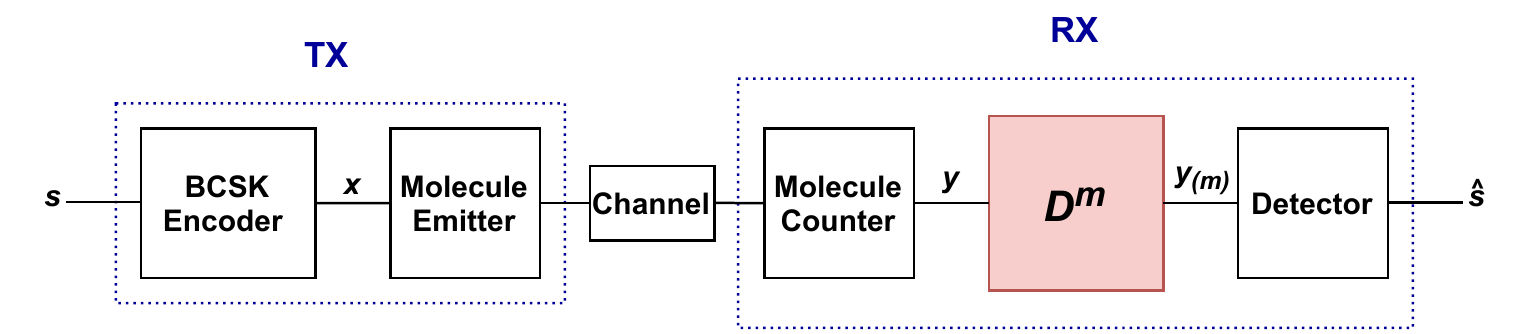} 
	\caption{Overall diagram of a derivative pre-processed MCD system.}
	\label{fig:systemdiagram}
	%\vspace*{-0.1in}
\end{figure*}

 Per its description in Section \ref{sec:proposed}, the $m^{th}$ order derivative operator can be interpreted as a pre-processor, whose output $\boldsymbol{y}_{(m)}$ is fed to the detector. Through this perspective, Figure \ref{fig:systemdiagram} presents the overall diagram of an end-to-end MCD system where the receiver employs the $\boldsymbol{D}^m$ operator. Herein, we address the design of the detector to be paired with the $m^{th}$ order derivative.

\subsection{Optimal Detector}

\par Our notion of optimality is defined by the maximum-likelihood criterion, which implies that the maximum likelihood sequence detector (MLSD) is optimal due to ISI in the MCD channel \cite{receiver_design}. Given that the receiver has access to $\lambda_s$ and $\boldsymbol{h}$, the MLSD estimates the transmitted bit sequence $\hat{\boldsymbol{s}}$ using the following rule:
\begin{equation} 
\label{eq:MLSD}
\begin{split}
\hat{\boldsymbol{s}} &= \argmax_{\boldsymbol{s}} P(\boldsymbol{y}_{(m)} | \boldsymbol{s})\\
&= \argmax_{\boldsymbol{s}} \frac{ \exp [-\frac{1}{2} (\boldsymbol{y}_{(m)} - \boldsymbol{\mu}_{(m)})^T \boldsymbol{\Sigma}^{-1}_{(m)} (\boldsymbol{y}_{(m)} - \boldsymbol{\mu}_{(m)}) ]}{{\sqrt{(2\pi)^{SN} |\boldsymbol{\Sigma_{(m)}}|}}} \\
&= \argmin_{\boldsymbol{s}} \Big\{ \ln(|\boldsymbol{\Sigma}_{(m)}|) + \\
& \hspace{2.0cm} (\boldsymbol{y}_{(m)} - \boldsymbol{\mu}_{(m)})^T \boldsymbol{\Sigma}^{-1}_{(m)} (\boldsymbol{y}_{(m)} - \boldsymbol{\mu}_{(m)}) \Big\}.
\end{split}
\end{equation}
Here, for each candidate symbol vector $\boldsymbol{s}$, the conditional $\boldsymbol{\mu}_{(m)}$ and $\boldsymbol{\Sigma}_{(m)}$ are found using the corresponding channel statistics presented in Equations \eqref{eq:matrix_form}-\eqref{eq:y_derivatived}. All vectors are of size $SN \times 1$ and the covariance matrix is $SN \times SN$.

\subsection{Banded MLSD}
\label{subsec:online_MLSD}

\par The complexity of the MLSD is exponential in channel memory $L$ using the Viterbi algorithm.
%as it computes and compares among $2^L$ candidate symbol sequences to detect a single symbol.
Unfortunately, as the data rate increases, a shorter bit duration $t_b$ implies a larger $L$ due to the heavy tail of the CIR, rendering MLSD infeasible for low-complexity, nano-scale machinery. However, leveraging the aggressive ISI mitigation introduced by the derivative operator, we consider a sub-optimal MLSD-like detector implemented using a considerably shorter memory window $L'$. This banded-MLSD approach results in a significantly lower complexity in computation, as it requires $2^{L'}$ log-likelihood computations to detect a single symbol, compared to $2^{L}$ required by MLSD (\cite{viterbi_older,viterbi_newer}). For this detector, the branch metric $\mathcal{M}(\cdot)$ that is input to the Viterbi decoder has the following form:
\begin{equation} \label{eq:online_MLSD}
\begin{split}
\mathcal{M}(\boldsymbol{y}^{(i)}_{(m)},\boldsymbol{s}^{(L')}) &= \ln(|\boldsymbol{\Sigma}^{(L')}_{(m)}|) + \\
& \hspace{0.5cm} (\boldsymbol{y}^{(i)}_{(m)}  - \boldsymbol{\mu}^{(L')}_{(m)})^\top {\boldsymbol{\Sigma}^{(L')}_{(m)}}^{-1} (\boldsymbol{y}^{(i)}_{(m)} - \boldsymbol{\mu}^{(L')}_{(m)}).
\end{split}
\end{equation}
Here, the superscript $(L')$ implies that the vectors and matrices are obtained considering a symbol memory of $L'$, and the superscript $(i)$ refers to the samples of the $i^{th}$ symbol. In particular, $\boldsymbol{s}^{(L')}$ denotes a candidate symbol string of length $L'$. Conditioned on a certain $\boldsymbol{s}^{(L')}$, $\boldsymbol{\mu}^{(L')}$ and $\boldsymbol{\Sigma}^{(L')}$ denote the obtained mean vector and covariance matrix of the samples of the $(L')^{th}$ symbol, respectively. Hence, the respective sizes of $\boldsymbol{\mu}^{(L')}$ and $\boldsymbol{\Sigma}^{(L')}$ are $N \times 1$ and $N \times N$. Furthermore, similar to its conventional use throughout the paper, the subscript $(m)$ implies that the argument vector is pre-multiplied by $\boldsymbol{D}^m$, and the argument matrix is pre- and post-multiplied by $\boldsymbol{D}^m$ and $(\boldsymbol{D}^{\top})^m$, respectively. Note that $\boldsymbol{D}^m$ is also of size $N \times N$ herein.

\par A key observation is that for a fixed derivative order $m$, $y_{(m)}[k]$ is a function of $y[k],\dots,y[k+m]$. Consequently, the $\boldsymbol{D}^m$ operator causes the last $m$ samples of the $i^{th}$ symbol to be correlated with the first samples of the $(i+1)^{th}$ symbol, which induces a non-causal ISI. To avoid this issue, we truncate the last $m$ samples of the intended symbol. Hence, $\boldsymbol{\mu}^{L'}_{(m)}$ and $\boldsymbol{\Sigma}^{L'}_{(m)}$ are of size $(N-m) \times 1$ and $(N-m) \times (N-m)$, respectively. Similarly, noting $\boldsymbol{y}^{(i)}$ corresponds to the arrival counts of the $i^{th}$ symbol, we have $\boldsymbol{y}^{(i)}_{(m)} = \begin{bmatrix} y_{(m)}[(i-1)N+1] & \dots & y_{(m)}[iN - m] \end{bmatrix}^\top$.

\par Assuming standard matrix multiplication, each branch metric computation has cubic complexity in the number of samples per symbol $N$, as the operation involves multiplying vectors and matrices of sizes that are linear in $N$. Note that $\boldsymbol{D}^m$ and the conditional vectors/matrices $\boldsymbol{\mu}^{(L')}_{(m)}$, $\boldsymbol{\Sigma}^{(L')}_{(m)}$, and ${\boldsymbol{\Sigma}^{(L')}_{(m)}}^{-1}$ can be pre-computed once and stored, which makes their complexity independent of $N$ and $m$. However, obtaining $\boldsymbol{y}^{(i)}_{(m)} $ from $\boldsymbol{y}^{(i)}$ is not independent of $m$. In particular, although the discrete time forward derivative operation is represented by $\boldsymbol{D}$ for clarity of argument, the operation for obtaining $\boldsymbol{y}^{(i)}_{(m)}$ can be realized with a simple shift register and element-wise subtractions. Therefore, the $m^{th}$ order derivative pre-processor has a complexity of $\mathcal{O}(mN)$ per symbol, making the overall complexity of banded MLSD $\mathcal{O}(m 2^{L'} N^3 )$.

\subsection{Decision Feedback-Aided, Symbol-by-Symbol ML} 

\par %As can be inferred from the banded-MLSD case, the aggressive ISI mitigation introduced by the $\boldsymbol{D}^m$ operator allows for a significantly shorter memory consideration in detector design.
The banded-MLSD's complexity is still exponential in $L'$, which might be undesirable for a nano-machine. To this end, we generalize the memory limited decision aided decoder (MLDA) proposed in \cite{MLDA_mitra_2014} for an arbitrary derivative order $m$, where $m=0$ corresponds to the original version of the detector.

\par In essence, MLDA is a decision feedback aided, symbol-by-symbol maximum likelihood detector. For each symbol, it first estimates the imposed ISI on the intended symbol's samples, using the previously decoded symbols. Ideally, the ISI estimation is done using all $L-1$ past decoded symbols. However, due to possible memory or computational constraints of a nano-machine, it might be desirable to consider a shorter memory window of $L'$ when estimating the ISI. For this memory-limited case, only $L'-1$ past decoded symbols are utilized,  and the rest of the past is replaced by the expected transmissions (\textit{i.e.,} using \eqref{eq:bcsk} with equiprobable transmissions). The entries of the estimated ISI mean of the $i^{th}$ symbol, denoted by $\hat{\boldsymbol{\mu}}^{(i)}_{\textrm{ISI}}$, is computed according to the previously decoded symbols as follows:
 \begin{equation} \label{eq:symbolML_past}
    \begin{split}
     	\hat{\mu}^{(i)}_{\textrm{ISI}}[n] = & \sum_{k=0}^{(L'-1)N - 1} h[k+n+1] \hspace*{0.1cm} \hat{x}[(i-1)N - k] \\
     	& + \sum_{k=(L'-1)N}^{(L-1)N-1} h[k+n+1] \hspace*{0.1cm} \bar{x}[(L-1)N - k],
    \end{split}
 \end{equation}
where $\boldsymbol{\hat{x}}$ denotes the decoded emission vector, and is computed through the decoded symbol vector $\boldsymbol{\hat{s}}$ through \eqref{eq:bcsk}. In addition, $\boldsymbol{\bar{x}}$ denotes the expected transmission vector that covers the past symbols between $(L'+1)^{th}$ and $L^{th}$ memory slots, and is expressed as
$$
\boldsymbol{\bar{x}} = \begin{bmatrix} \frac{M}{2} & 0 & \dots & 0 & \frac{M}{2} & 0 & \dots \end{bmatrix}^\top_{(L-L')N \times 1}.
$$
After estimating the ISI-induced mean vector, estimated distributions for the samples of possible bit-1 and bit-0 transmissions are computed. Let the estimated (and Gaussian approximated) arrival random vectors be denoted as $\hat{\boldsymbol{y}}^{(i)}_{1} \sim \mathcal{N}(\hat{\boldsymbol{\mu}}^{(i)}_{1},\hat{\boldsymbol{\Sigma}}^{(i)}_{1})$ and $\hat{\boldsymbol{y}}^{(i)}_{0} \sim \mathcal{N}(\hat{\boldsymbol{\mu}}^{(i)}_{0},\hat{\boldsymbol{\Sigma}}^{(i)}_{0})$. The $i^{th}$ sample's estimated means can be written as 
 \begin{equation} 
 \label{eq:MLDA_estimate}
 	\begin{split}
 		\hat{\mu}^{(i)}_{1}[n] &= \hat{\mu}^{(i)}_{\textrm{ISI}}[n] + M h[n] + \lambda_s , \\
 		\hat{\mu}^{(i)}_{0}[n] &= \hat{\mu}^{(i)}_{\textrm{ISI}}[n] + \lambda_s,
 	\end{split}
 \end{equation}
where, $\hat{\boldsymbol{\Sigma}}^{(i)}_{j} = \textrm{diag} \{\hat{\boldsymbol{\mu}}^{(i)}_{j}\}$ for $j \in \{0, 1\}$, as $\hat{\boldsymbol{y}}^{(i)}_{j}$ are Gaussian approximations of Poisson RVs. 

\par Recalling the truncated arrival vector output of the derivative operator from Subsection \ref{subsec:online_MLSD} as $$\boldsymbol{y}^{(i)}_{(m)} = \begin{bmatrix} y_{(m)}[(i-1)N+1] & \dots & y_{(m)}[iN - m] \end{bmatrix}^\top,$$ the symbol is detected through a likelihood ratio test, that is
\begin{equation}
\hat{s}[i] = \hat{\mathcal{L}}^{(i)}_{(m),1} - \hat{\mathcal{L}}^{(i)}_{(m),0} \mathop{\gtrless}_0^1 0,
\end{equation}
where the log-likelihoods are computed as 
\begin{equation}
	\label{eq:loglik_DFEML}
	\begin{split}
    	\hat{\mathcal{L}}^{(i)}_{(m),j} = -&\frac{1}{2} \Big[ (N-m)\ln(2\pi) + \ln (|\hat{\boldsymbol{\Sigma}}^{(i)}_{(m),j})|) + \\
    	&(\boldsymbol{y}^{(i)}_{(m)} - \hat{\boldsymbol{\mu}}^{(i)}_{(m),j})^\top \left(\hat{\boldsymbol{\Sigma}}^{(i)}_{(m),j}\right)^{-1} (\boldsymbol{y}^{(i)}_{(m)} - \hat{\boldsymbol{\mu}}^{(i)}_{(m),j}) \Big]. 
	\end{split}
\end{equation}
Here, $\hat{\boldsymbol{\mu}}^{(i)}_{(m),j} = \boldsymbol{D}^m \hat{\boldsymbol{\mu}}^{(i)}_{j}$ and $\hat{\boldsymbol{\Sigma}}^{(i)}_{(m),j} = \boldsymbol{D}^m \hat{\boldsymbol{\Sigma}}^{(i)}_{j} (\boldsymbol{D}^m)^\top$.

\par Overall, MLDA provides a computationally cheaper alternative to banded MLSD, by incurring a linear computational complexity in $L'$. Note that for each symbol, evaluating Equations \eqref{eq:symbolML_past}-\eqref{eq:MLDA_estimate} necessitates holding $(L'-1)$ past symbols, and \eqref{eq:symbolML_past} involves the element-wise multiplication of two $(L'-1) N$ sample-long vectors. Including the $\mathcal{O}(mN)$ complexity of the $m^{th}$ order differentiation and the $\mathcal{O}(N^3)$ of \eqref{eq:loglik_DFEML}, the complexity of MLDA is of order $\mathcal{O}(m L' N^3)$.

\subsection{Fixed Threshold Detectors}
\label{subsec:FTDs}

\par Up to this point, each considered detector employs memory at the receiver side. However, low-complexity and memoryless detectors are particularly desirable for nano-scale applications. To this end, we consider two types of fixed threshold detectors in this subsection. Both detectors rely on comparing the arrival count at a certain sample with a threshold, but they differ in their selection of the arrival count to be compared.

\subsubsection{Max-and-Threshold Detector}

\par The max-and-threshold detector (MaTD) selects the sample with the maximum arrival count among the samples corresponding to the intended symbol \cite{lin_derivative,gursoymitra_globecom}. The detection steps of the $\boldsymbol{D}^m$-MaTD pair can be summarized as follows:
\begin{itemize}
    \item Employ the derivative operator,
    \item Discard the last $m$ samples (to cancel non-causal ISI),
    \item Perform an $\argmax$ operation among the remaining $(N-m)$ samples, 
    \item Compare with the threshold.
\end{itemize}
In essence, the detected symbol is found by performing
\begin{equation} \label{eq:_max_threshold_detector}
	\hat{s}[i] = \max \big(y_{(m)}[(i-1)N+1],\cdots,y_{(m)}[iN-m] \big) \mathop{\gtrless}_0^1 \gamma,
\end{equation}
where $\gamma$ is the employed fixed threshold. We note that MaTD with $m=0$ corresponds to the \textit{simple asynchronous detector} (ADS) proposed in \cite{ADDF_ADS}.

\subsubsection{Fixed Sample, Fixed Threshold Detector}
\label{subsubsec:FSTD}

\par Using MaTD, the sample that observes the maximum number of molecules may differ for each transmitted symbol, as the arrival counts are stochastic. In contrast, as we considered in our prior study \cite{gursoymitra_ACM}, the threshold detector can also be realized by fixing the sample whose arrival count is to be compared, yielding the \textit{fixed sample, fixed threshold detector} (FSTD). In particular, denoting the fixed sample of interest as $\tilde{q}_{(m)}$, FSTD selects $\tilde{q}_{(m)}$ as the ``peak sample due to the intended symbol". In other words, $\tilde{q}_{(m)}$ corresponds to the sample that has the largest expected arrival count due to the intended symbol's transmission. Let $\boldsymbol{\mu}_{s,(m)}$ denote the expected signal due to the intended symbol after the $m^{th}$ order derivative operator is applied at the receiver. We note that since the peak of $\boldsymbol{\mu}_{s,(m)}$ changes with $m$ (see Figure \ref{fig:fhits}), $\tilde{q}_{(m)}$ is a function of $m$. Overall, the $\boldsymbol{\mu}_{s,(m)}$ vector can be expressed as
\begin{equation}
    \label{eq:mu_sm}
    \boldsymbol{\mu}_{s,(m)} = \boldsymbol{D}^m \boldsymbol{\mu}_{s},
\end{equation}
where
\begin{equation}
\label{eq:intended_mean} 
\boldsymbol{\mu}_{s} = 
    \begin{cases}
      \begin{bmatrix}
            M h[1] & \dots & M h[N]
            \end{bmatrix}^{\top}, & \text{if bit-$1$} \\
      \begin{bmatrix}
            0 & \dots & 0
            \end{bmatrix}^{\top}, & \text{if bit-$0$}.
    \end{cases}
\end{equation}
Similar to previously discussed strategies, for a derivative order of $m$, the last $m$ samples of $\boldsymbol{\mu}_{s,(m)}$ are again discarded to avoid non-causal ISI. Following this truncation, and denoting $\boldsymbol{\bar{\mu}}_{s,(m)}$ as the expected $\boldsymbol{\mu}_{s,(m)}$, FSTD selects $\tilde{q}_{(m)}$ by performing
\begin{equation}
\label{eq:tilde_q}
\tilde{q}_{(m)} = \argmax_{q \in \{1,\dots,N-m\}} \hspace{0.3cm} \left|\bar{\mu}_{s,(m)}[q] \right|.
\end{equation}
The maximization in \eqref{eq:tilde_q} does not perform the $\argmax$ operation on $\boldsymbol{\mu}_{s,(m)}$ itself, but selects the sample with the largest signal in the absolute sense. Note that due to the nature of time differentiation and the $f_{hit}(t)$ function, $\boldsymbol{\mu}_{s,(m)}$ can have both positive and negative elements (see Figure \ref{fig:fhits}). In some cases, the smallest negative element can actually have a larger absolute value than the largest positive element, implying that said negative sample is larger in energy. In such a case, FSTD simply negates the received signal and finds $\tilde{q}_{(m)}$ using the negated signal. Overall, the decision rule for FSTD can be expressed as 
\begin{equation} \label{eq:fixedsample_threshold_detector}
\hat{s}[i] = 
\begin{cases}
    y_{(m)}[(i-1)N + \tilde{q}_{(m)}] \mathop{\gtrless}_0^1 \gamma, & \textrm{if} \hspace{0.15cm} {\mu}_{s,(m)}[\tilde{q}_{(m)}] \geq 0 \\
    -y_{(m)}[(i-1)N + \tilde{q}_{(m)}] \mathop{\gtrless}_0^1 \gamma, & \textrm{if} \hspace{0.15cm} {\mu}_{s,(m)}[\tilde{q}_{(m)}] < 0.
\end{cases}
\end{equation}
We note that FSTD is a generalization of the fixed sample, fixed threshold detector that is widely used in the MCD literature (\cite{ftd_1,ftd_2,ftd_3}) to an arbitrary derivative order $m$, where $m=0$ corresponds to the original version of the detector.

\par Since both MaTD and FSTD are memoryless detectors, their complexities for decoding a symbol do not depend on a memory window length $L'$. For FSTD, as the threshold comparison is done using a fixed sample, the complexity does not depend on $N$ either. For MaTD, finding the maximum on $N-m$ samples has linear complexity in $N$. Overall, combining with the $\mathcal{O}(mN)$ of the derivative pre-processing, both MaTD and FSTD have complexities of order $\mathcal{O}(mN)$. This result suggests the conjunction of $\boldsymbol{D}^m$ and fixed threshold detectors are particularly useful for low complexity, nano- to micro-scale applications. Motivated by this, we will mainly consider fixed threshold detectors throughout the rest of the paper, with a particular focus of FSTD for the problem of derivative order optimization. In the numerical results, Section \ref{sec:results}, we will compare performance of all of the detectors discussed herein.

\section{The Optimization of $m$}
\label{sec:costfunction}

\subsection{Error Probability Analysis}
\label{subsec:BER}

\par With the derivative order $m$ as a design parameter, the question of how to optimize it arises. Herein, we address this derivative order optimization problem for fixed threshold detectors. As the end goal is to minimize the error rate of the transmission, we first derive the theoretical bit error probability of the $\boldsymbol{D}^m$-FSTD pair. 

\par Recalling that the considered MCD channel is an ISI channel with memory length $L$, the theoretical error probability expression will average over all $(L-1)$ symbol-long strings of data. Denoting $\boldsymbol{s}_{\textrm{ISI}}$ as the $(L-1)$ symbol-long vector that holds said string, the error probability can be found by performing
\begin{equation} 
\label{eq:Pe_overall}
P_{e} = \frac{1}{2^{L-1}} \big(\sum_{\forall \boldsymbol{s}_{\textrm{ISI}}} P_{e|\boldsymbol{s}_{\textrm{ISI}}}\big).
\end{equation}
Conditioned on a certain symbol vector $\boldsymbol{s}_L = \begin{bmatrix} \boldsymbol{s}_{\textrm{ISI}} & s[L] \end{bmatrix}^{\top}$, the received signal mean can be written as
\begin{equation}
\label{eq:hypo_for_bound}
\begin{split}
\boldsymbol{\mu}_L &= E(\boldsymbol{y}_L|\boldsymbol{x}_L) \\
&= \boldsymbol{H}_{L} \boldsymbol{x}_L + \lambda_s \boldsymbol{j}_N,
\end{split}
\end{equation}
where $\boldsymbol{j}_N$ is an $N$ sample-long vector of ones, the received vector $\boldsymbol{y}_L = \begin{bmatrix}
y[(L-1)N+1] & \cdots & y[LN] 
\end{bmatrix}^T$, $\boldsymbol{x}_L$ is the corresponding $LN$ sample-long transmission vector corresponding to $\boldsymbol{s}_L$ through \eqref{eq:bcsk}, and 
\begin{equation}
\label{eq:H_for_bound}
\boldsymbol{H}_L = \begin{bmatrix}
h[(L-1)N+1] & \cdots &  h[1] & 0 & \cdots & 0 \\
\vdots & \cdots &   & \ddots & \ddots & \vdots \\
h[L N-1] & \cdots &   & h[2] & h[1] & 0 \\
h[L N] &  \cdots &  & h[3] & h[2] & h[1]
\end{bmatrix}.
\end{equation}
Similar to \eqref{eq:matrix_form2}, the covariance matrix is then found by $\boldsymbol{\Sigma}_{L} = \textrm{diag}\{\boldsymbol{\mu}_{L}\}$. Therefore, after applying the $m^{th}$ order derivative operator, the mean vector and covariance matrix associated with each conditional becomes $\boldsymbol{\mu}_{L,(m)} = \boldsymbol{D}^m \boldsymbol{\mu}_{L}$ and $\boldsymbol{\Sigma}_{L,(m)} = \boldsymbol{D}^m \boldsymbol{\Sigma}_{L} (\boldsymbol{D}^{\top})^m$, respectively.

\par Using these conditional statistics, we are interested in finding $P_{e|\boldsymbol{s}_{\textrm{ISI}}}$, which can be expressed for FSTD as
\begin{equation}
\label{eq:conditional_general}
\begin{split}
P_{e|\boldsymbol{s}_{\textrm{ISI}}} = &\frac{1}{2} \big(P(B \hspace{0.05cm} y_{(m)}[(L-1)N+\tilde{q}_{(m)}] < \gamma | s_{L}[L] = 1) \\ &+ P(B \hspace{0.05cm} y_{(m)}[(L-1)N+\tilde{q}_{(m)}] > \gamma | s_{L}[L] = 0) \big) \\
= &\frac{1}{2} (A_1 + A_0),
\end{split}
\end{equation}
where $\tilde{q}_{(m)}$ is the fixed sample found by Equations \eqref{eq:intended_mean}-\eqref{eq:tilde_q}, and $B = \operatorname{sgn} \left( \mu_{s,(m)}[\tilde{q}_{(m)}] \right)$ with $\operatorname{sgn}(\cdot)$ defining the signum function. As we employ the Gaussian approximation of the Poisson arrivals, \eqref{eq:conditional_general} can be re-written as
\begin{equation}
\label{eq:conditional_qfunc}
	\begin{split}
		A_1 = Q \Big( \frac{B \hspace{0.05cm} \mu_{L,(m)}[\tilde{q}_{(m)}] - \gamma}{\sqrt{\Sigma_{L,(m)}[\tilde{q}_{(m)},\tilde{q}_{(m)}]}} \Big| s_L[L] = 1 \Big) \\
		A_0 = Q \Big( \frac{\gamma - B \hspace{0.05cm} \mu_{L,(m)}[\tilde{q}_{(m)}] }{\sqrt{\Sigma_{L,(m)}[\tilde{q}_{(m)},\tilde{q}_{(m)}]}} \Big| s_L[L] = 0 \Big),
	\end{split}
\end{equation}
where $Q(\cdot)$ is the Gaussian $Q$-function, which concludes the derivation. Note that $A_1$ and $A_0$ consider the same $\boldsymbol{s}_{\textrm{ISI}}$. However, they differ in $s_L[L]$, hence the mean vectors and covariance matrices presented in \eqref{eq:conditional_qfunc} are not equal.

\par \textbf{Error Analysis of the $\boldsymbol{D}^m$-MaTD Pair:} As noted in Subsection \ref{subsec:FTDs}, we focus on FSTD as the primary threshold-based detector in this paper. However, we also provide the error probability derivation of the $\boldsymbol{D}^m$-MaTD pair, as it may be more desirable in scenarios where the receiver is not capable of locating the expected signal peak location for FSTD.

\par Similar to FSTD, due to the ISI nature of the MCD channel, the error probability of $\boldsymbol{D}^m$-MaTD is also found by averaging over the conditional error probabilities. Similarly, the computation of conditional statistics is also not dependent on the detector strategy. Therefore, Equations \eqref{eq:Pe_overall}-\eqref{eq:H_for_bound} hold for the $\boldsymbol{D}^m$-MaTD pair as well. 

\par The derivation for the $\boldsymbol{D}^m$-MaTD pair differs from that of the $\boldsymbol{D}^m$-FSTD pair in the way it computes the conditional error probabilities. To evaluate a conditional error probability for the $\boldsymbol{D}^m$-MaTD pair, we first denote $$Y_{(m)} = \underset{j \in \{1,\dots,N-m\}}\max \hspace{0.05cm}y_{L,(m)}[j]$$ as the maximum sample. Then, $P_{e|\boldsymbol{s}_{\textrm{ISI}}}$ can be expressed as
\begin{equation}
\label{eq:cond_matd}
\begin{split}
    P_{e|\boldsymbol{s}_{\textrm{ISI}}} &= \frac{1}{2} \big(P(Y_{(m)} < \gamma | s_{L}[L] = 1)
    + P(Y_{(m)} > \gamma | s_{L}[L] = 0) \big) \\
    &= \frac{1}{2} (A'_1 + A'_0)
\end{split}
\end{equation}
Therefore, characterizing the CDF of $Y_{(m)}$ is sufficient to complete the derivation. However, $Y_{(m)}$ corresponds to the maximum of correlated and differently distributed Gaussian random variables, for which a straightforward, closed form solution does not appear to exist. Instead, numerical solutions are typically considered \cite{corr_gaus2,corr_gaus}. Motivated by this, we use the strategy developed by Clark \cite{ClarkApprox} to approximate $Y_{(m)}$ as a normal random variable through a recursive process. We refer the reader to Appendix \ref{ap:clark} for the details of the recursion. At the end of the Clark's approximation process, we obtain the approximate Gaussian distribution $Y_{(m)} \sim \mathcal{N}(\mu_R,\sigma^2_R)$. We plug these statistics into \eqref{eq:cond_matd} as
\begin{equation}
\label{eq:conditional_qfunc_matd}
	\begin{split}
		A'_1 = Q \Big( \frac{\mu_R - \gamma}{\sigma_R }\Big| s_L[L] = 1 \Big) \\
		A'_0 = Q \Big( \frac{\gamma - \mu_R}{\sigma_R } \Big| s_L[L] = 0 \Big),
	\end{split}
\end{equation}
which completes the derivation.

\subsection{Signal-to-Interference-Plus-Noise Ratio}
\label{subsec:SINR}

\par 
%As it is the true objective function, optimizing the derivative order $m$ through evaluating the theoretical bit error probability for each $m$ yields an accurate optimization strategy. However, carrying out the derivations presented in Subsection \ref{subsec:BER} necessitate evaluating $2^{L-1}$ conditional error probabilities, which is exponential in $L$ and may incur an infeasibly high computational complexity since $L$ is typically very large for MCD systems. 

Due to the presence of the $2^{L-1}$ conditional error probabilities which constitute the theoretical BER, optimizing the BER resulting from a particular choice of $m$ is computationally infeasible given the typical values of $L$ associated with an MCD channel.
In particular, even though our earlier study suggested considering $L' < L$ to decrease this complexity \cite{gursoymitra_globecom}, our observations suggest that such a simplification can lose accuracy for very high data rate settings, even with the aggressive right tail mitigation of the $\boldsymbol{D}^m$ operator. Furthermore, as for the case of true $L$, we still need to compute multiple $Q$-functions for $L'$ as well (Equation \eqref{eq:conditional_qfunc}), which may be
%we note that each evaluation of \eqref{eq:conditional_qfunc} involves computing $Q$-function values, which may be 
undesirable for a simple nano-machine.

\par Motivated by the aforementioned shortcomings of optimizing $m$ by examining the theoretical error probabilities, we generalize the signal-to-interference-plus-noise ratio (SINR) employed by \cite{jamali_mf} to the $\boldsymbol{D}^m$-FSTD pair. Overall, for an arbitrary symbol index $k$, the expression has the following form \cite{gursoymitra_ACM}:
\begin{equation}
\label{eq:SINR}
\begin{aligned} 
\operatorname{SINR}(m) &=\frac{E\left\{\left( s[k] \mu_{s,(m)}[\tilde{q}_{(m)}] \right)^{2}\right\}}{\operatorname{Var}\left\{ \eta_{(m)}^{(k)} [\tilde{q}_{(m)}]\right\}+\operatorname{Var}\left\{ \mathcal{I}_{(m)}^{(k)}[\tilde{q}_{(m)}] \right\}}, %\\
\end{aligned}
\end{equation}
where the superscript $(k)$ indicates that the symbol of interest is the $k^{th}$. In the sequel, we characterize each term in the expression. Firstly, the numerator corresponds to the second moment of the signal that is induced by the intended symbol's ($k^{th}$) transmission. Recalling the definition of $\boldsymbol{\mu}_{s,(m)}$ from \eqref{eq:mu_sm} and that $\tilde{q}_{(m)}$ is $\boldsymbol{D}^m$-FSTD's sample of interest, the numerator is expressed as 
\begin{equation}
    \label{eq:nominator}
	\begin{split}
		E\left\{ \left( s[k] \mu_{s,(m)}[\tilde{q}_{(m)}] \right)^{2} \right\} &= \frac{1}{2} \times 0 + \frac{1}{2} E\left\{ \left( \mu_{s,(m)}[\tilde{q}_{(m)}] \right)^{2} \right\} \\
		&= \frac{1}{2} \left( \mu_{s,(m)}[\tilde{q}_{(m)}] \right)^{2}.
	\end{split}
\end{equation}
In the denominator, the first expression represents the noise variance induced by the intended symbol's transmission. Let $\boldsymbol{\Sigma}_{s} = \operatorname{diag}(\boldsymbol{\mu}_{s})$ and $\boldsymbol{\Sigma}_{s,(m)} = \boldsymbol{D}^m \boldsymbol{\Sigma}_{s} (\boldsymbol{D}^m)^\top$. Then, the noise variance incurred by the intended symbol is expressed as
\begin{equation}
	\begin{split}
		\operatorname{Var}\left\{ \eta_{(m)}^{(k)} [\tilde{q}] \right\} &= \frac{1}{2} \operatorname{Var}\left\{ \eta_{(m)}^{(k)} [\tilde{q}] \Big| s[k] = 0 \right\} + \\ & \hspace{2cm} \frac{1}{2} \operatorname{Var}\left\{ \eta_{(m)}^{(k)} [\tilde{q}] \Big| s[k] = 1 \right\} \\
		&= \frac{1}{2} \times 0 + \frac{1}{2} \boldsymbol{\Sigma}_{s,(m)} [\tilde{q}_{(m)},\tilde{q}_{(m)}] \\
	    &= \frac{1}{2} \left\{\boldsymbol{D}^m \operatorname{diag}(\boldsymbol{\mu}_{s})  (\boldsymbol{D}^m)^{\top} \right\} [\tilde{q}_{(m)},\tilde{q}_{(m)}].
	\end{split}
\end{equation}
Lastly, characterizing the noise variance induced by ISI and external noise completes the derivation of \eqref{eq:SINR}. To this end, we first denote $\boldsymbol{\mu_{I}}$ as the mean arrival count vector that is due to ISI. Note that $\boldsymbol{\mu_{I}}$ depends on the evaluated ISI symbol sequence $\boldsymbol{s}_{\textrm{ISI}}$. We then define
\begin{equation}
\begin{split}
    \boldsymbol{\mu}_{I,(m)} &= \boldsymbol{D}^m \left( \boldsymbol{\mu_{I}} + \lambda_s \boldsymbol{j} \right) \\
    \boldsymbol{\Sigma}_{I,(m)} &= \boldsymbol{D}^m \operatorname{diag}(\boldsymbol{\mu_{I}} + \lambda_s \boldsymbol{j}) (\boldsymbol{D}^{\top})^m.
\end{split}
\end{equation}
%Note that $\boldsymbol{\mu}_{I,(0)} \neq \boldsymbol{\mu_{I}}$, since $\boldsymbol{\mu}_{I,(0)}$ also includes the external noise effect within its definition. 
Overall, the variance induced by ISI and external noise can be found by 
\begin{equation}
    \label{eq:ISI_var}
	\begin{split}
		\operatorname{Var}& \left\{ \mathcal{I}_{(m)}^{(k)}[\tilde{q}_{(m)}] \right\} = \operatorname{Cov}(y_{(m)}^{(k)}[\tilde{q}_{(m)}],y_{(m)}^{(k)}[\tilde{q}_{(m)}]) \\
		&= E\left[ \left(y_{(m)}^{(k)}[\tilde{q}_{(m)}] \right)^2\right] - \left( E \left[ y_{(m)}^{(k)}[\tilde{q}_{(m)} ]\right] \right)^2  \\
		&= E_{\boldsymbol{s}_{ISI}}\left[ \left(y_{(m)}^{(k)}[\tilde{q}_{(m)}] \right)^2 \Big| \boldsymbol{s}_{ISI} \right] - \\
		& \hspace{3.5cm} \left( E_{\boldsymbol{s}_{ISI}} \left[ y_{(m)}^{(k)}[\tilde{q}_{(m)} \big| \boldsymbol{s}_{ISI}]\right] \right)^2  \\
		&= \frac{1}{2^{L-1} } \sum_{\forall \boldsymbol{s}_{ISI}}^{ } \left[\boldsymbol{\Sigma}_{I,(m)}[\tilde{q}_{(m)},\tilde{q}_{(m)}] + \left(\mu_{I,(m)}[\tilde{q}_{(m)}]\right)^2 \right] - \\
		& \hspace{3.5cm} \left(\frac{1}{2^{L-1} } \sum_{\forall \boldsymbol{s}_{ISI}}^{ } \mu_{I,(m)}[\tilde{q}_{(m)}] \right)^2.
	\end{split}
\end{equation}

\par Similar to the theoretical BER expressions, evaluating SINR also incurs an exponential complexity in $L$ stemming from computing conditional statistics when evaluating \eqref{eq:ISI_var}. To avoid this, one can use a smaller memory window of $L' < L$  when evaluating SINR, significantly reducing the incurred complexity. Denoting this version of SINR as $\operatorname{SINR}_{L'}$, the objective function presented in this subsection can be used to select the derivative order as follows:
\begin{equation}
    m^* = \argmax_{m} \hspace{0.15cm} \operatorname{SINR}_{L'}(m).
\end{equation}

\section{Numerical Results}
\label{sec:results}

\par In this section, we present numerical results to assess the accuracies of the theoretical error probability expressions derived in Subsection \ref{subsec:BER}, demonstrate the efficacy of SINR as an objective function for optimizing $m$, and provide comparative BER results for derivative/no-derivative detectors. Throughout the section, the external Poisson noise rate ($\lambda_s$) is normalized with respect to transmission power through the following definition of signal-to-noise ratio (SNR):
\begin{equation}
    \label{eq:SNR}
    \textrm{SNR} = \frac{\frac{M}{2}}{N \lambda_s} = \frac{M}{2 N\lambda_s}.
\end{equation}
Here, the numerator follows from \eqref{eq:bcsk} for equiprobable BCSK symbols, and represents the average emitted signal per one symbol. Recalling $N$ as the number of samples per one symbol duration and $\lambda_s$ as the external noise rate per symbol, the denominator of \eqref{eq:SNR} represents the expected number of external noise molecules per one symbol. 

\par Since parameters such as $D$, $r_r$, and $r_0$ all affect the $f_{hit}(t)$ function hence the $\boldsymbol{h}$ vector. To this end, in order to contextualize the data rate in relation to these parameters, we normalize the symbol duration with respect to the channel peak time $t_p = \frac{d^2}{6D} = \frac{(r_0-r_r)^2}{6D}$, see \cite[Equation 26]{3Dchar}. Throughout this section, the bit duration $t_b$ is selected through a unitless parameter $S_r$, which is defined as $S_r = \frac{t_b}{t_p}$. Note that since a smaller $t_b$ corresponds to a higher rate of transmission, a smaller $S_r$ corresponds to a higher data rate.

\begin{figure*}[!t]
	\centering
	\subfloat[$S_r = 0.5$.]{\label{fig:theo_05}\includegraphics[width=.48\textwidth]{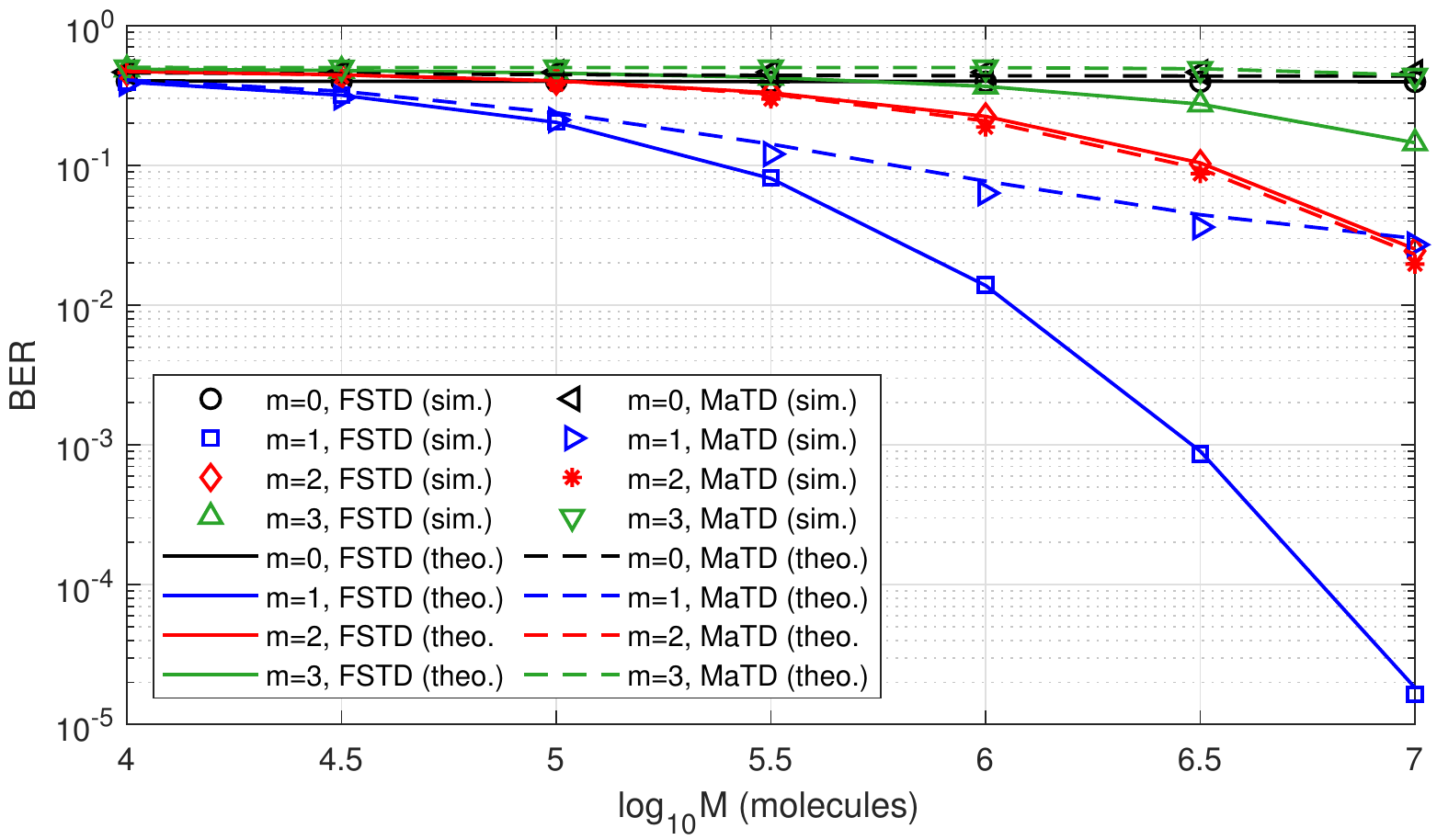}}
    \subfloat[$S_r = 0.25$.]{\label{fig:theo_025}\includegraphics[width=.48\textwidth]{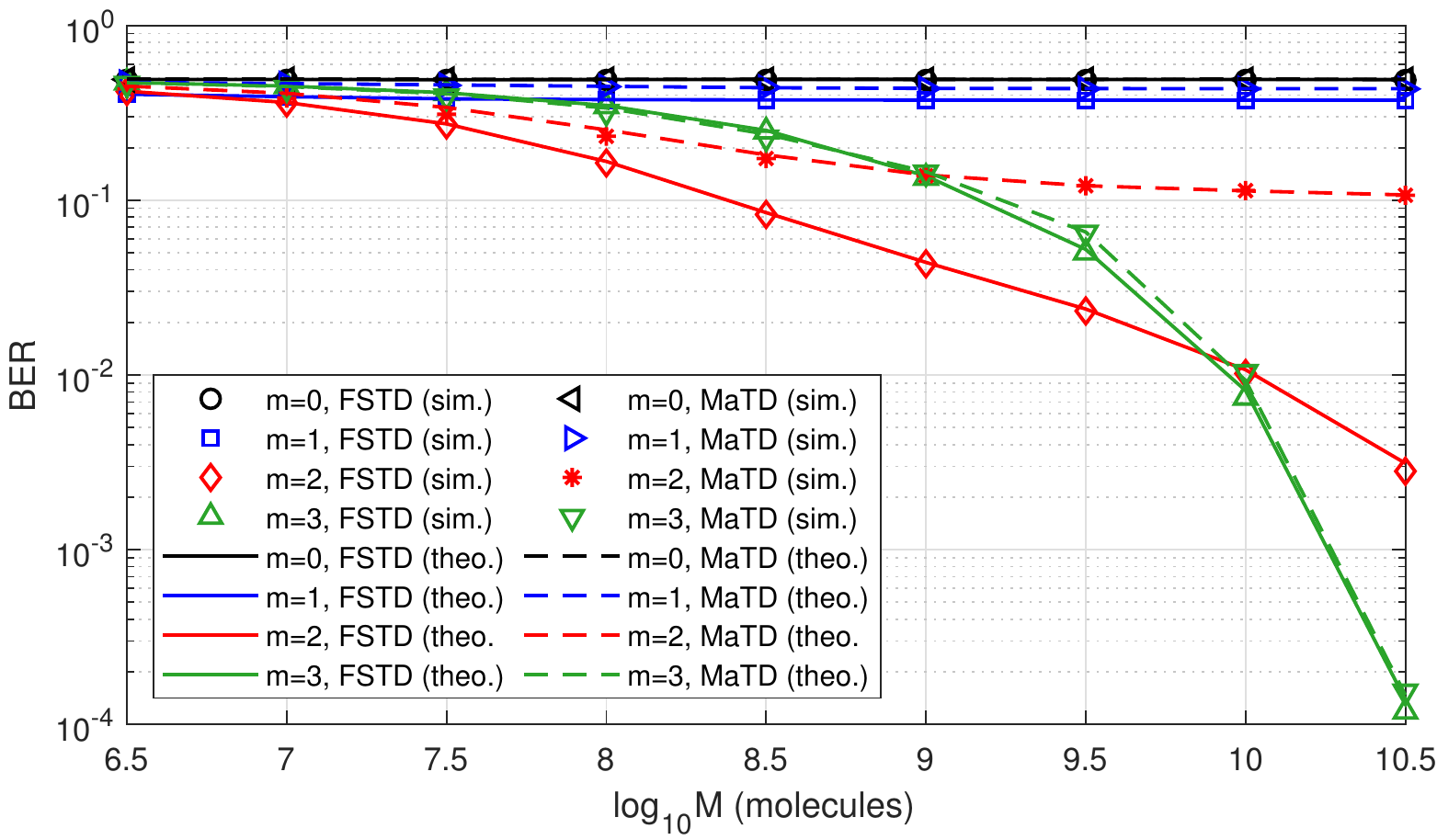}}
	\caption{Theoretical and simulated BER vs. $M$. $\textrm{SNR} = 10$dB, $r_0 = \SI{15}{\micro \meter}$, $r_r = \SI{5}{\micro \meter}$, $D = \SI{100}{\micro \meter \squared \per\second}$, $N=5$, $L=10$. $\gamma$ values numerically optimized through exhaustive search.}
	\label{fig:BER_theo}
\end{figure*}

\subsection{Accuracy of Error Analysis}

\par Herein, we demonstrate the accuracy of the derived theoretical BER expressions for $\boldsymbol{D}^m$-FSTD and $\boldsymbol{D}^m$-MaTD pairs with varying $m$. To this end, Figures \ref{fig:theo_05} and \ref{fig:theo_025} are presented for different values of $S_r$ (hence, different data rates). In both figures, $L=10$ is selected for demonstrative purposes, due to the exponential complexity when computing the theoretical BER expressions.

% \begin{figure}[!t]
% 	\centering
% 	\includegraphics[width=.48\textwidth]{Sr05_theoexp_FSTD_MaTD.eps}
% 	\caption{BER vs. $M$. $S_r = 0.5$, $\textrm{SNR} = 10$dB, $r_0 = \SI{15}{\micro \meter}$, $r_r = \SI{5}{\micro \meter}$, $D = \SI{100}{\micro \meter \squared \per\second}$, $N=5$, $L=10$. $\gamma$ values numerically optimized through exhaustive search.}
% 	\label{fig:theo_05}
% \end{figure}

% \begin{figure}[!t]
% 	\centering
% 	\includegraphics[width=.48\textwidth]{Sr025_theoexp_FSTD_MaTD.eps}
% 	\caption{BER vs. $M$. $S_r = 0.25$, $\textrm{SNR} = 10$dB, $r_0 = \SI{15}{\micro \meter}$, $r_r = \SI{5}{\micro \meter}$, $D = \SI{100}{\micro \meter \squared \per\second}$, $N=5$, $L=10$. $\gamma$ values numerically optimized through exhaustive search.}
% 	\label{fig:theo_025}
% \end{figure}

\par The results of Figures \ref{fig:theo_05} and \ref{fig:theo_025} show that the theoretical BER expression for $\boldsymbol{D}^m$-FSTD is accurate, and the approximations for $\boldsymbol{D}^m$-MaTD are tight. Furthermore, confirming the results of \cite{gursoymitra_ACM}, FSTD is found to generally outperform MaTD. Motivated by this, among the fixed threshold detectors, we will present the error curves for $\boldsymbol{D}^m$-FSTD throughout this section. Lastly, regardless of the comparative relationship between FSTD and MaTD, it can be observed that both detectors benefit from the derivative operator and produce lower BER values with $m > 0$ compared to their standard versions with $m=0$.

\subsection{Accuracy of SINR}
\label{subsec:results_SINR}

\par
\begin{figure*}[!t]
	\centering
	\subfloat[SINR vs. $M$, $S_r = 0.5$, $L'=10$.]{\label{fig:SINR_05}\includegraphics[width=.48\textwidth]{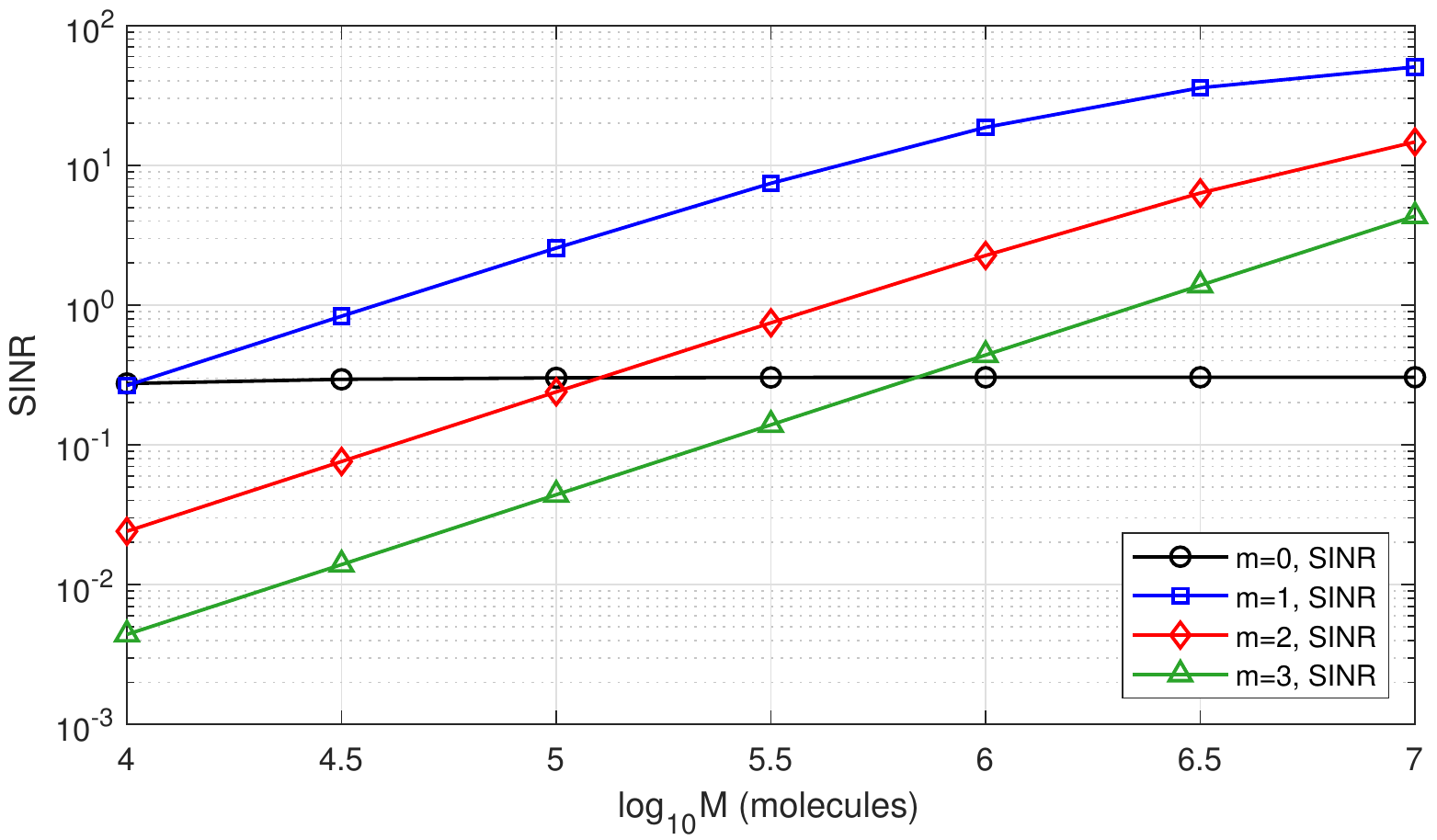}}
	\subfloat[Theoretical and Simulated BER vs. $M$, $S_r = 0.5$, $L=100$.]{\label{fig:FSTD_05}\includegraphics[width=.48\textwidth]{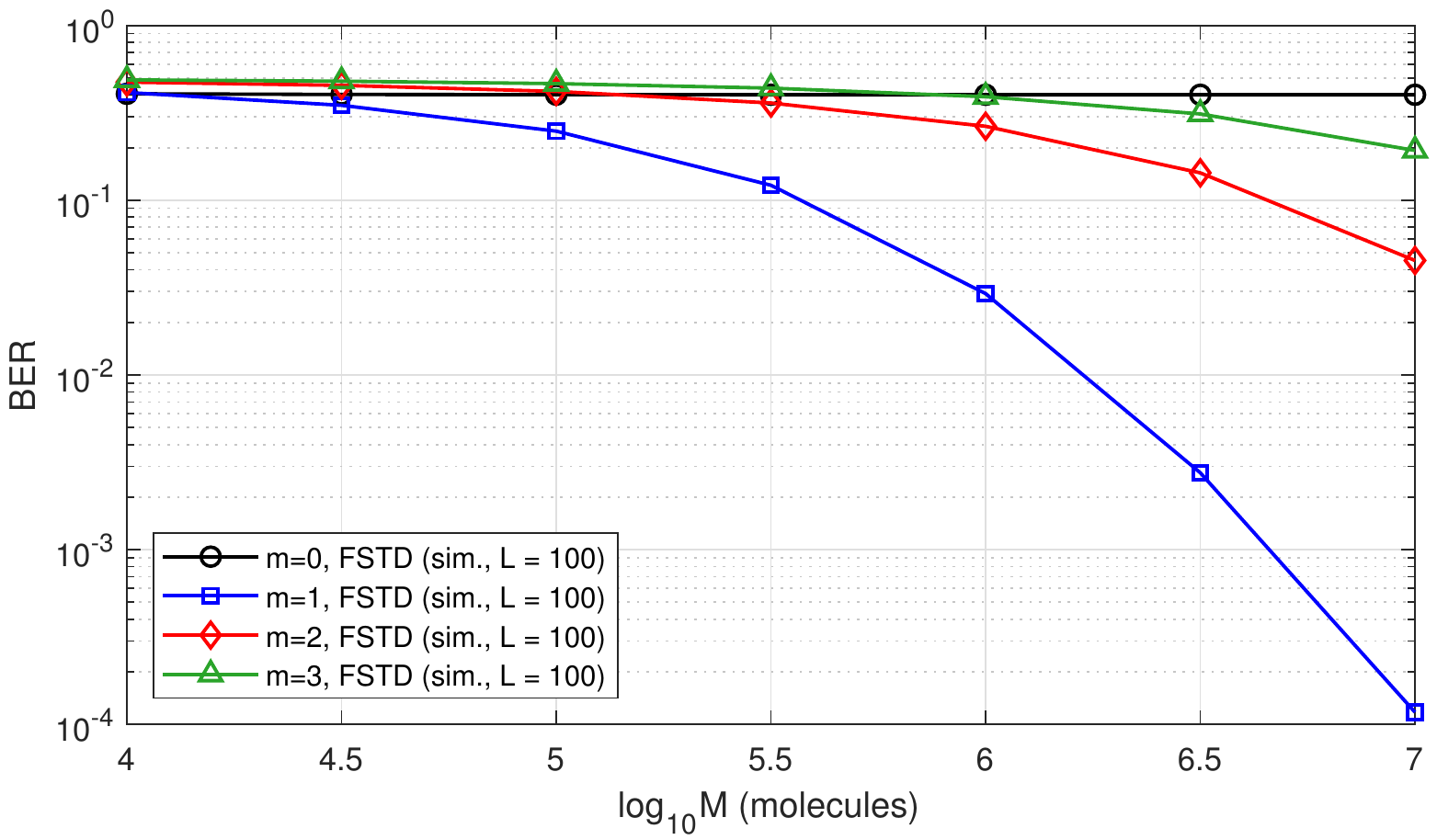}} \quad
    \subfloat[SINR vs. $M$, $S_r = 0.25$, $L'=10$.]{\label{fig:SINR_025}\includegraphics[width=.48\textwidth]{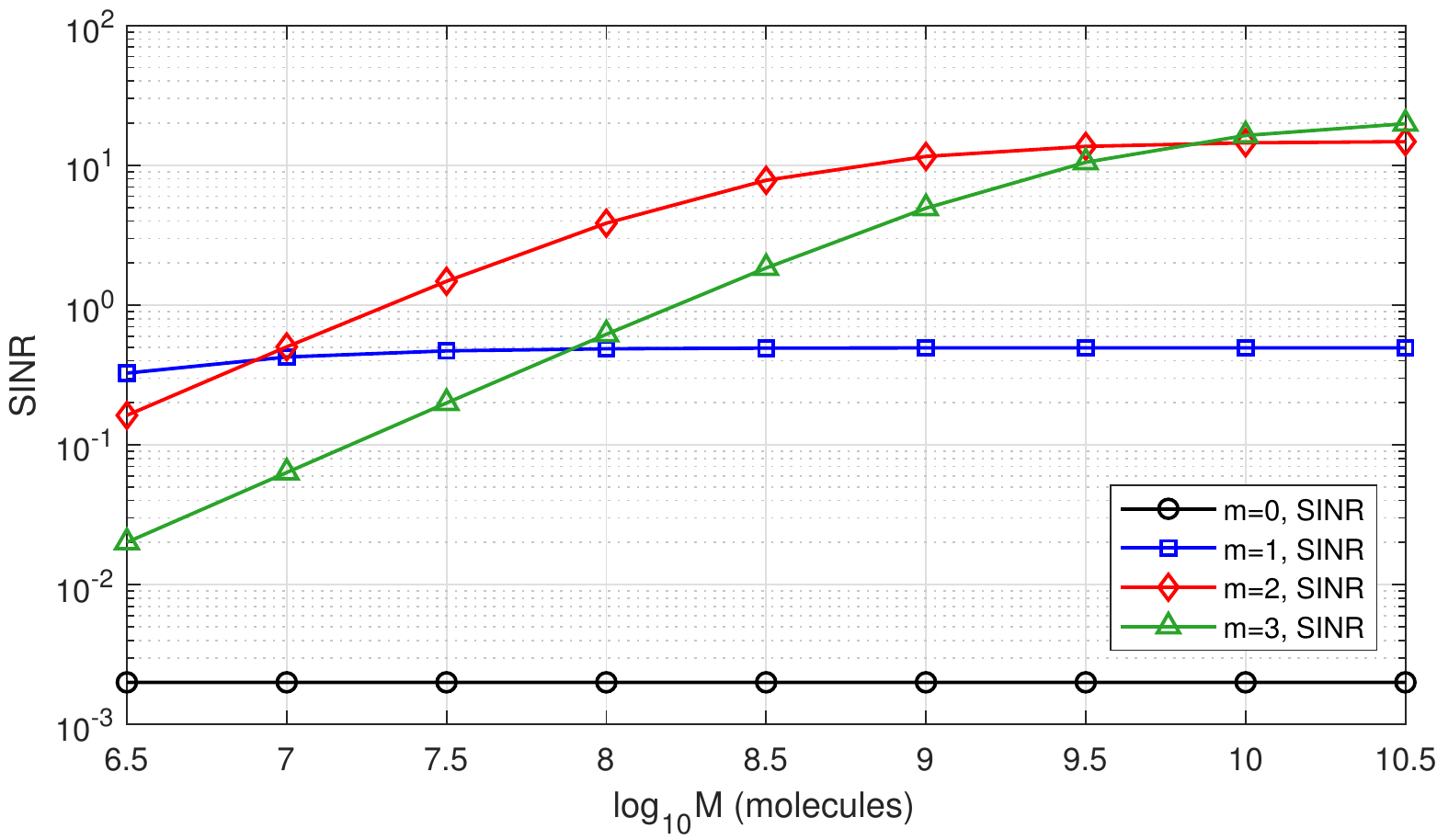}} 
    \subfloat[Theoretical and Simulated BER vs. $M$, $S_r = 0.25$, $L=200$.]{\label{fig:FSTD_025}\includegraphics[width=.48\textwidth]{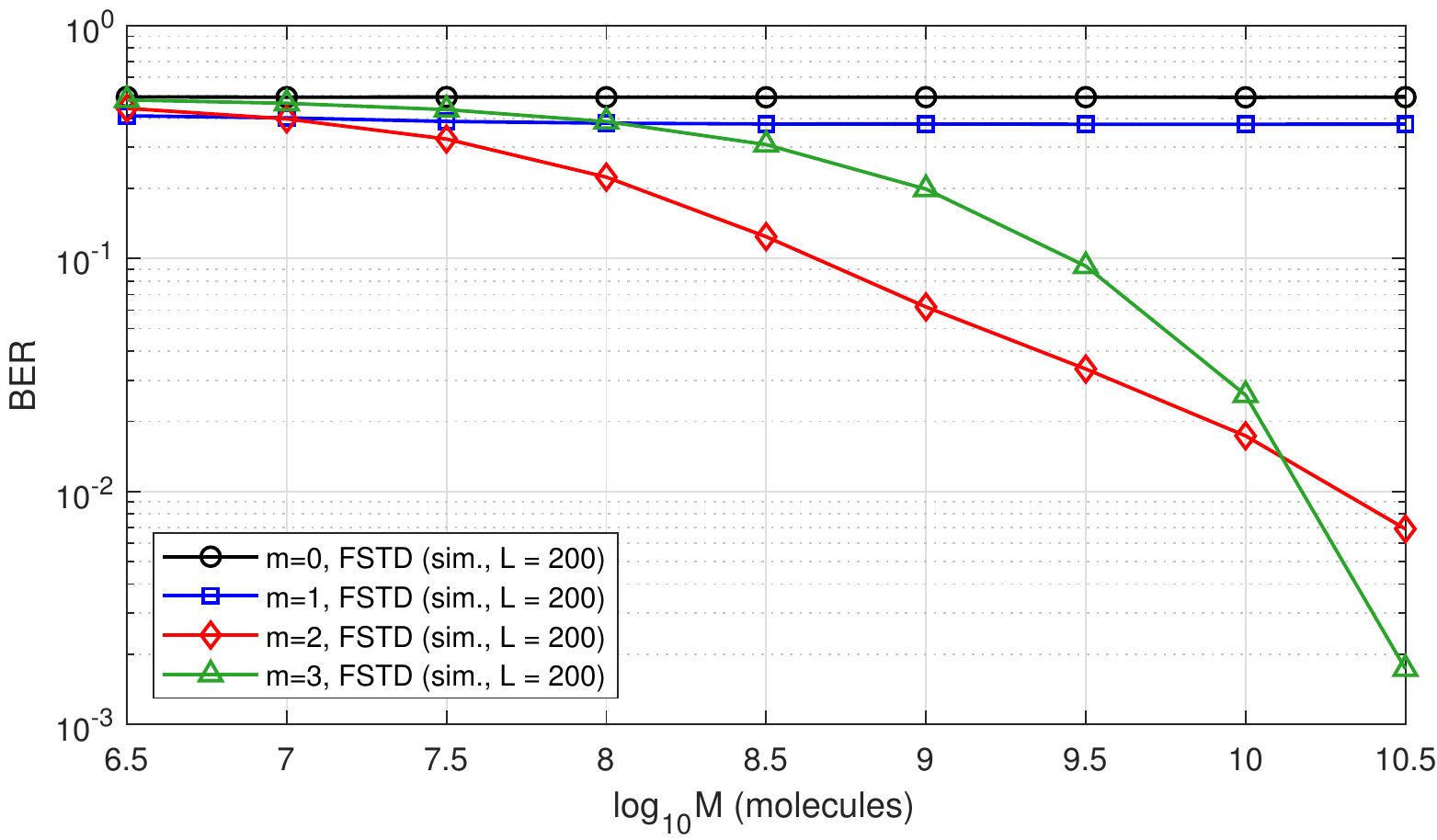}} \\    
	\caption{BER and SINR vs. $M$. $\textrm{SNR} = 10$dB, $r_0 = \SI{15}{\micro \meter}$, $r_r = \SI{5}{\micro \meter}$, $D = \SI{100}{\micro \meter \squared \per\second}$, $N=5$. $\gamma$ values numerically optimized through exhaustive search.}
	\label{fig:SINR_FSTD}
\end{figure*}

 In this subsection, we show the accuracy of the SINR expression derived in Subsection \ref{subsec:SINR}. To this end, we provide Figures \ref{fig:SINR_05}-\ref{fig:FSTD_05} and Figures \ref{fig:SINR_025}-\ref{fig:FSTD_025} to present results for two different data rates, and thus, two different levels of ISI. Similar to the theoretical error probability expressions, the SINR also necessitates evaluating over $2^{L-1}$ ISI symbol sequences. For computational complexity reasons, we use the limited memory, $\operatorname{SINR}_{L'}$ version of the expression with $L' = 10$. However, we note that the BER simulations use large channel memories ($L=100$ for Figure \ref{fig:FSTD_05} and $L=200$ for Figure \ref{fig:FSTD_025}) to satisfactorily capture the right tail of the CIR, and to test the memory-limited SINR's efficacy in the more accurate large channel memory scenario.

\par The results of Figure \ref{fig:SINR_FSTD} demonstrate that SINR closely follows the comparative trend between different derivative orders. Furthermore, SINR provides this accuracy with a significantly smaller memory consideration than the true channel memory, which suggests its utility for derivative order optimization in micro- to nano-scale applications. That said, SINR can incur slight discrepancies in the comparative trend when the BER values of evaluated schemes are close. An example of this phenomenon can be observed in Figures \ref{fig:SINR_025}-\ref{fig:FSTD_025}, between $m=2$ and $3$ at $M = 10^{10}$ molecules. The discrepancy is due to the substantially smaller memory used to compute the SINR. We refer the reader to compare Figures \ref{fig:SINR_025} and \ref{fig:theo_025}, which are with $L = L' = 10$, to confirm SINR's accuracy when the considered memories are equal.

\par The comparative trends between different orders of $m$ in Figures \ref{fig:FSTD_05} and \ref{fig:FSTD_025} show that, as theorized and predicted, the optimal derivative order is a function of system parameters. In particular, Figure \ref{fig:FSTD_05} shows that for a relatively smaller $S_r$ (hence lower ISI) a smaller $m$ is better. On the other hand, Figure \ref{fig:FSTD_025} shows that in a larger $S_r$/higher ISI regime, higher derivative orders outperform the first order. These results can be explained through the fundamental trade-off between ISI mitigation and noise amplification associated with the $\boldsymbol{D}^m$ operator. Recall from Section \ref{sec:proposed} and Figure \ref{fig:fhits} that a higher derivative order induces a lower ISI due to a narrower effective pulse duration, at a cost of an increase in received signal variance. In light of this trade-off, the results of Figure \ref{fig:FSTD_05} show that for lower ISI, the system is better off by avoiding the additional noise amplification of $m>1$ as the ISI is already relatively low\footnote{However, it should be noted that the optimal derivative order is still larger than $m=0$, indicating that the existing ISI is significant. We emphasize that $S_r = 0.5$ implies the bit duration is half of that of the channel peak duration, which still incurs a highly deteriorating level of ISI, hence $m>0$ is needed for meaningful communication.}. However, the higher data rate in Figure \ref{fig:FSTD_025} incurs a very high level of ISI, which induces the need for a more aggressive ISI mitigation, causing the optimal $m$ to be larger than one. Another noteworthy trend in Figure \ref{fig:FSTD_025} is that the optimal derivative order changes with $M$. Figure \ref{fig:FSTD_025} shows that in the small $M$ regime, $m=2$ is optimal. However, as $M$ increases, the system is able to combat the noise amplification better, hence is able to leverage the more powerful ISI mitigation provided by $m=3$.

\subsection{Asymptotic Performance}

\par
\begin{figure*}[!t]
	\centering
	\subfloat[$S_r = 0.5$.]{\label{fig:Sr05_SINRasymp}\includegraphics[width=.48\textwidth]{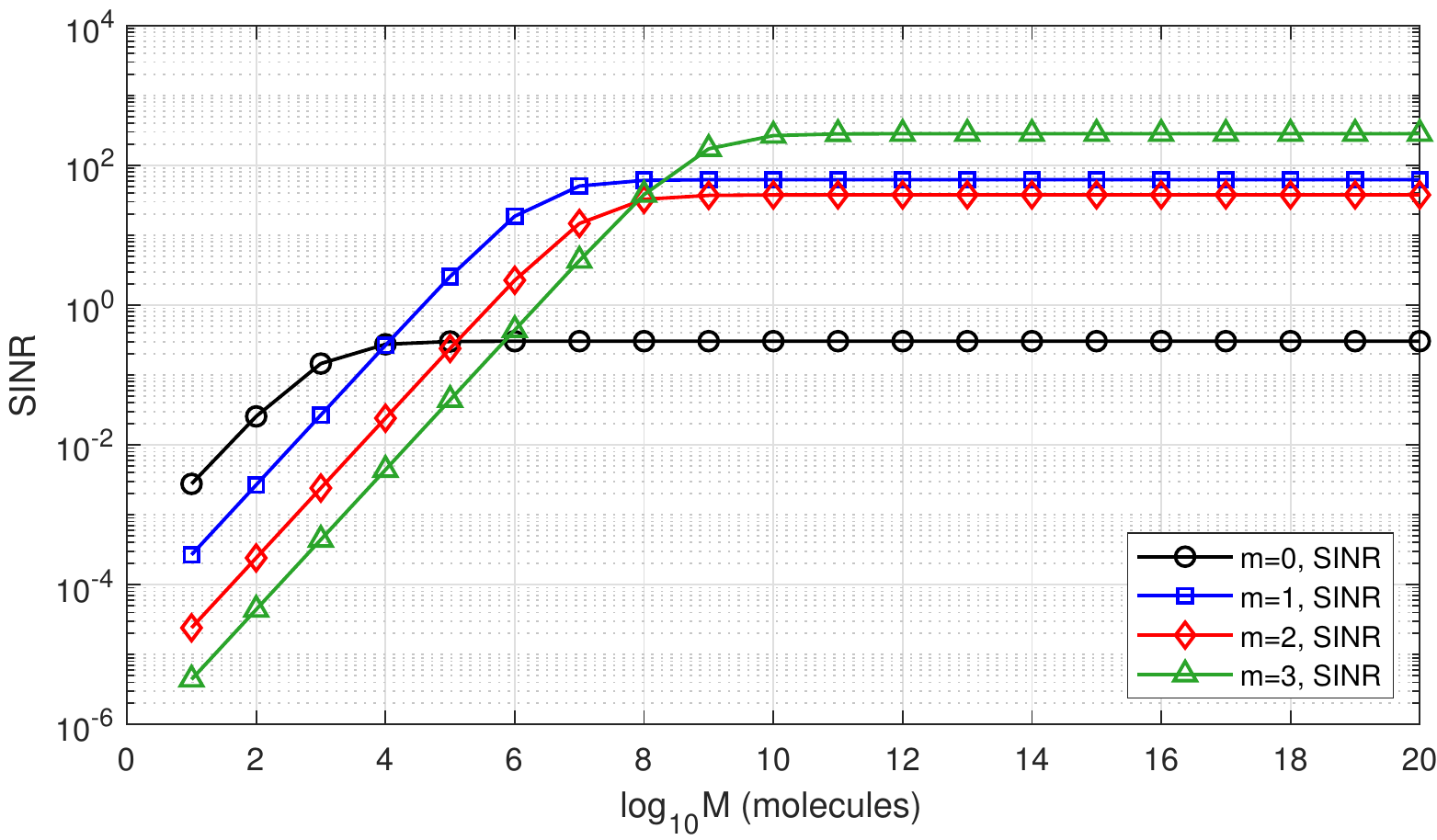}}
    \subfloat[$S_r = 0.25$.]{\label{fig:Sr025_SINRasymp}\includegraphics[width=.48\textwidth]{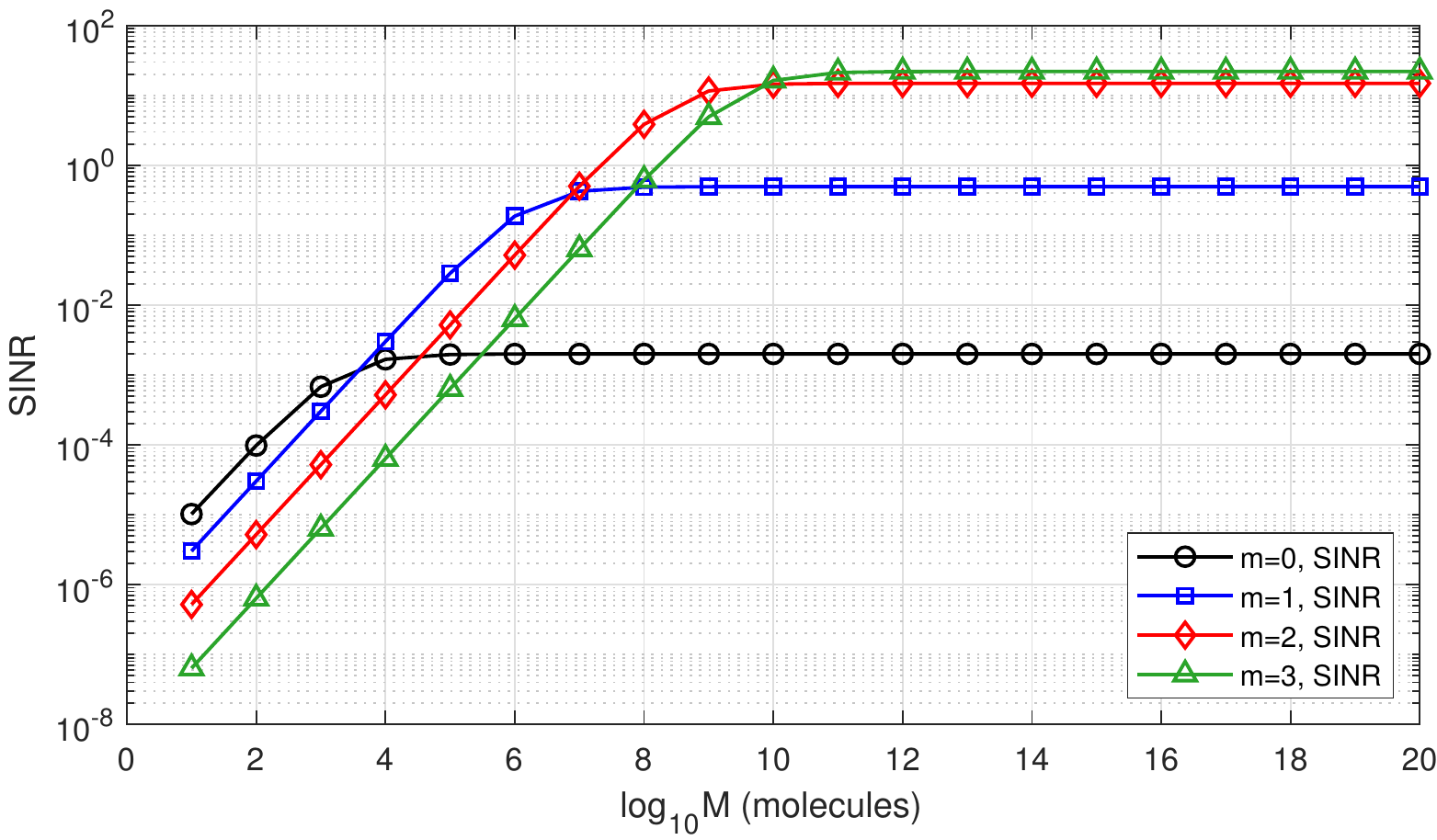}}
	\caption{SINR vs. $M$. $\textrm{SNR} = 10$dB, $r_0 = \SI{15}{\micro \meter}$, $r_r = \SI{5}{\micro \meter}$, $D = \SI{100}{\micro \meter \squared \per\second}$, $N=5$, $L'=10$.}
	\label{fig:SINR_asymp}
\end{figure*}

 In the previous subsection, we discussed the implications of the ISI mitigation-noise amplification trade-off of the derivative pre-processor. In particular, we noted that for low ISI and/or small $M$ scenarios, a smaller $m$ is better due to less noise amplification. On the other hand, in general, as $M$ increases, the system becomes more robust against noise, and is better-off by increasing the derivative order for better ISI mitigation. From these two trends, the following question might arise: As $M \rightarrow \infty$, do the performances of derivative orders become monotonically better as $m$ increases? To this end, although BER vs. $M$ curves cannot be provided due to extremely low error rates, we leverage the accuracy of SINR in explaining the comparative relationship of different derivative orders for $\boldsymbol{D}^m$-FSTD, and provide SINR vs. $M$ curves in Figure \ref{fig:SINR_asymp}.

\par As expected, the results of Figure \ref{fig:Sr025_SINRasymp} show that the SINR is monotonically decreasing in $m$ at the small $M$ regime, whereas for asymptotically large $M$, the SINR is monotonically increasing. Our empirical observations with various channel and system parameters verify that this trend is typical, confirming the implications of the ISI mitigation-noise amplification trade-off. However, Figure \ref{fig:Sr05_SINRasymp} exemplifies that it is not \textit{always} the case. We return to the sampling argument for Equation \eqref{eq:tilde_q}. In the context of \eqref{eq:tilde_q}, we had noted that time differentiation causes some samples of the derivative pre-processed signal to have large negative amplitudes in expectation. In addition to this, for some scenarios, the same effect of time differentiation can also cause the samples to more \textit{evenly} share the total received power within a symbol duration. In such cases, the maximum sample considered by FSTD has a smaller absolute magnitude, making the system face a higher error floor due to ISI. The computation of SINR implicitly accounts for this phenomenon and predicts the comparative relationships for asymptotically large $M$.

\subsection{BER vs. Detector Memory}

\par
\begin{figure}[!t]
	\centering
	\includegraphics[width=.48\textwidth]{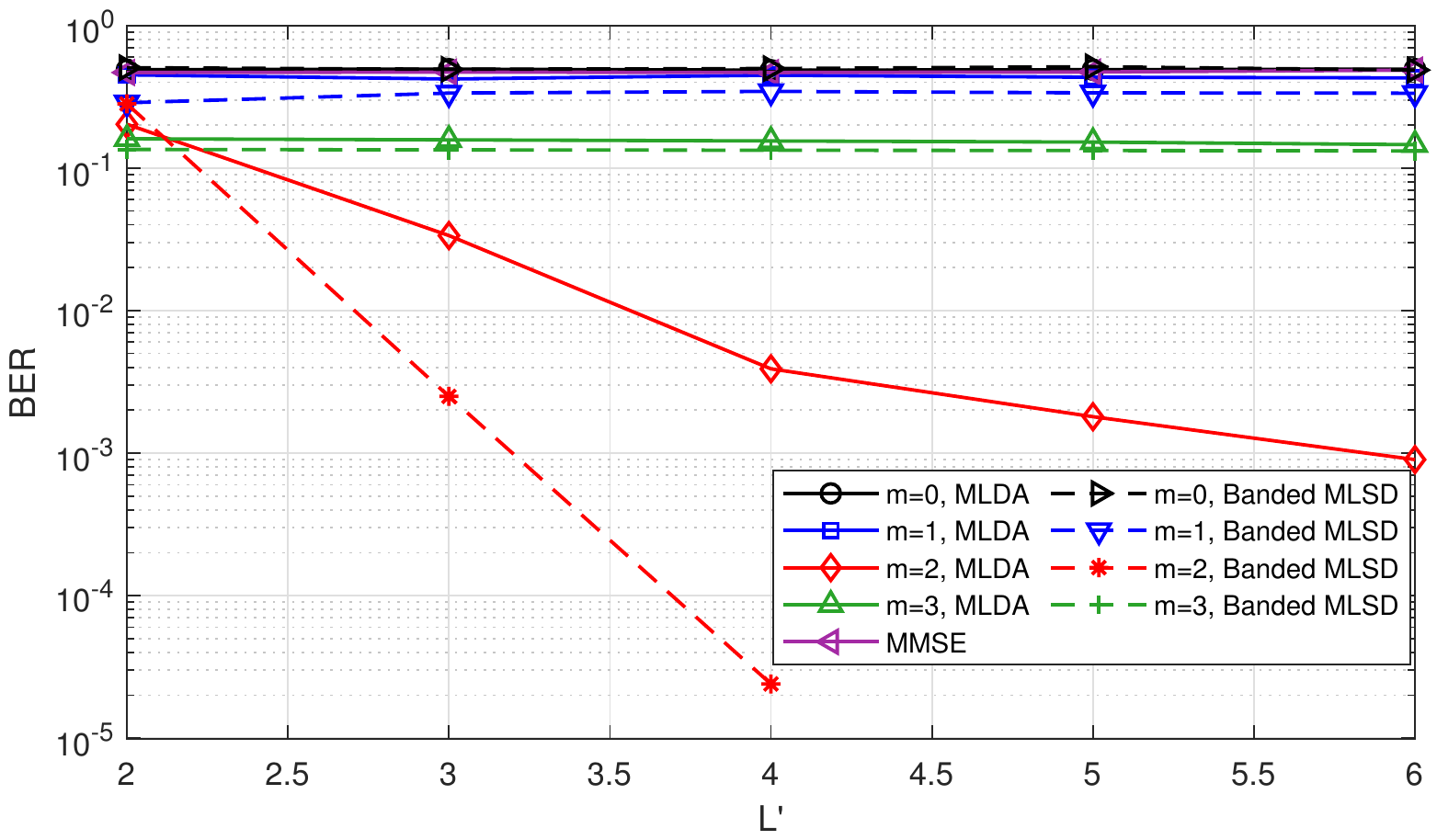}
	\caption{BER vs. $L'$. $S_r = 0.25$, $\log_{10}M = 8.5$ molecules, $\textrm{SNR} = 10$dB, $r_0 = \SI{15}{\micro \meter}$, $r_r = \SI{5}{\micro \meter}$, $D = \SI{100}{\micro \meter \squared \per\second}$, $N=5$, $L=200$.}
	\label{fig:BERvsL_025}
\end{figure}

 We next consider the effects of detector memory in error performance for the banded MLSD and MLDA. Figure \ref{fig:BERvsL_025} presents BER versus $L'$. For benchmarking purposes, Figure \ref{fig:BERvsL_025} also includes the results for minimum mean squared error (MMSE) equalizer. We note that the MMSE, when applied on the transmission block as a whole, has complexity that is quadratic in block length $S$, which is undesirable for a simple nano-machine. Hence, following the consideration of \cite{receiver_design}, we employ the decision feedback-aided, online version of the MMSE equalizer herein (which is $\mathcal{O}(L'^2 N^2)$ for the emission strategy considered in Equation \eqref{eq:bcsk}).

\par Figure \ref{fig:BERvsL_025} demonstrates three noteworthy trends:
\begin{enumerate}
    \item Derivative-based pre-processing allows for a reduction in the memory window of memory-aided detectors, courtesy of its ISI mitigating nature. We emphasize that even though the true channel memories are on the order of hundreds of symbols, reliable communication can be achieved using a substantially smaller $L'$. Combined with the very low complexity nature of the discrete-time derivative operation itself, the combination of the pre-processor and detector remain low complexity, providing high-performance, computationally cheap MCD receivers amenable to micro- and nano-scale MCD applications. 
    
    \item The ISI mitigation-noise amplification trade-off affects the optimal derivative order in memory-aided detectors as well. However, in memory-aided detectors, the optimal derivative order is also affected by the choice of $L'$. To exemplify, we note that the stronger ISI mitigation of $m=3$ makes the schemes with $m=3$ outperform other orders at $L'=2$. On the other hand, the schemes with $m=2$ provide lower error rates for $L'\geq3$ due to less noise amplification.
    
    \item Due to its very definition, as $L' \rightarrow L$, the performance of the banded MLSD would converge to the MLSD Viterbi decoder. For the same derivative order, the results of Figure \ref{fig:BERvsL_025} show that banded MLSD typically outperforms its MLDA counterpart in the small $L'$ regime as well. That said, we note that MLDA is also capable of yielding a reliable error performance at this regime, and is a low complexity alternative to banded MLSD therein. Overall, we conclude that $\boldsymbol{D}^m$-banded MLSD is able to provide a lower BER with higher complexity, and vice versa for $\boldsymbol{D}^m$-MLDA, confirming the performance-complexity trade-off between them.
\end{enumerate}

\begin{figure*}[!t]
	\centering
	\subfloat[BER vs. $M$, $S_r = 0.5$, $L=100$.]{\label{fig:Sr05_Lprime2}\includegraphics[width=.48\textwidth]{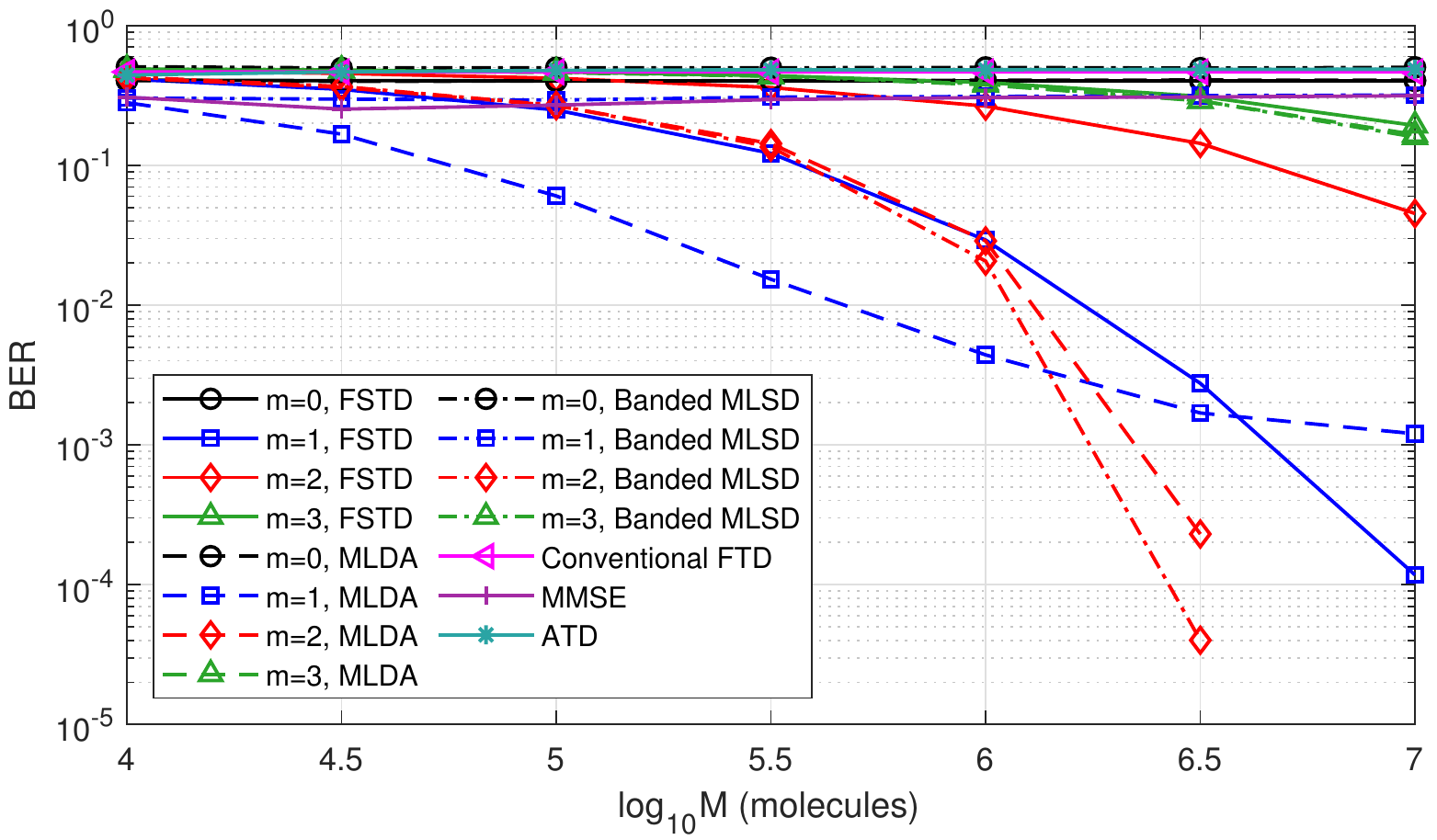}}
% 	\subfloat[BER vs. $M$, $S_r = 0.5$, $L'=5$, $L=100$.]{\label{fig:Sr05_Lprime5}\includegraphics[width=.48\textwidth]{Sr05_Lprime5_all.eps}} \quad
    \subfloat[BER vs. $M$, $S_r = 0.25$, $L=200$.]{\label{fig:Sr025_Lprime2}\includegraphics[width=.48\textwidth]{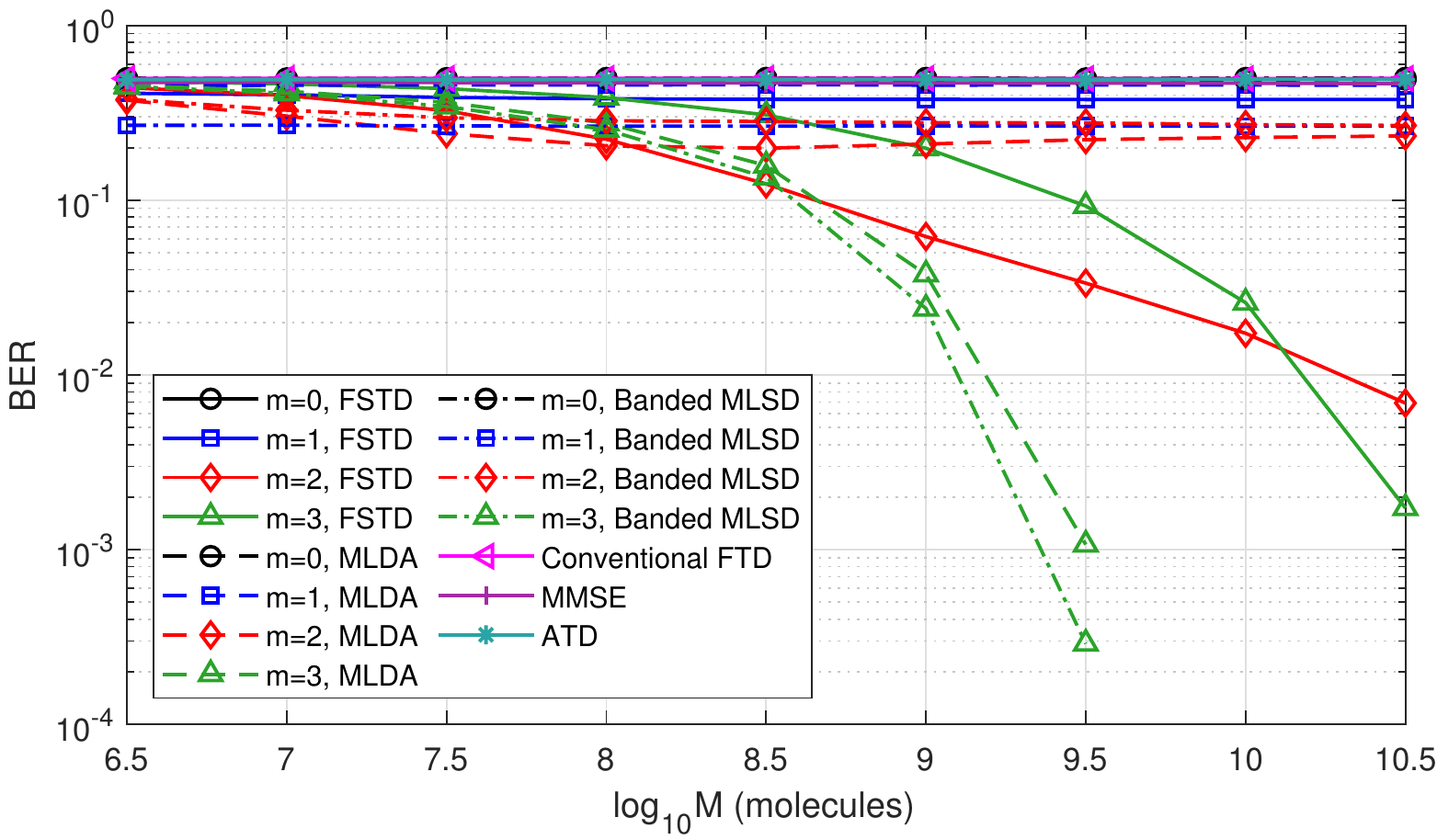}} 
    % \subfloat[BER vs. $M$, $S_r = 0.25$, $L'=5$, $L=200$.]{\label{fig:Sr025_Lprime5}\includegraphics[width=.48\textwidth]{Sr025_Lprime5_all.eps}} \\    
	\caption{BER vs. $M$. $\textrm{SNR} = 10$dB, $r_0 = \SI{15}{\micro \meter}$, $r_r = \SI{5}{\micro \meter}$, $D = \SI{100}{\micro \meter \squared \per\second}$, $N=5$, $L'=2$. $\gamma$ values numerically optimized through exhaustive search.}
	\label{fig:BER_comparative}
\end{figure*}

\subsection{Comparative Error Performance}

\par In this subsection, we provide the comparative error performances of the proposed strategies through Figure \ref{fig:BER_comparative}. We note that due to their comparable computational complexities, Figure \ref{fig:BER_comparative} also includes the conventional fixed threshold detector (FTD, \cite{CSKMOSK}) where the total arrival count within a symbol duration is compared with a threshold, and the adaptive threshold detector (ATD) proposed in \cite{ATD_2016}. Moreover, in order to ensure a relatively comparable complexity to FSTD, the memory aided detectors/equalizers are implemented with $L'=2$.

\par The results of Figure \ref{fig:BER_comparative} agree with the previously presented results regarding the benefits of the derivative operator. Among the derivative-based schemes, for the majority of the evaluated data points in Figure \ref{fig:BER_comparative}, either $\boldsymbol{D}^m$-banded MLSD or $\boldsymbol{D}^m$-MLDA are observed to outperform other schemes. That said, $\boldsymbol{D}^m$-FSTD is also found to provide reliable error rates despite its simplicity. Comparing the strategies on a target BER level, it can be inferred that $\boldsymbol{D}^m$-banded MLSD and $\boldsymbol{D}^m$-MLDA are able to reach the target with a smaller $M$ than $\boldsymbol{D}^m$-FSTD, whereas $\boldsymbol{D}^m$-FSTD offers a lower receiver complexity at a cost of increased transmission power.

\section{Conclusions}
\label{sec:conclusion}

\par In this paper, receiver-side higher order time differentiation has been proposed to mitigate ISI for MCD. Considering the derivative operation as a pre-processor block before detection, several memory-aided (\textit{i.e.}, MLSD, banded MLSD, MLDA) and memoryless (\textit{i.e.}, FSTD, MaTD) detectors have been provided to be paired with the derivative operator. In the paper, it is shown that for a derivative-based MCD receiver, there exists a fundamental trade-off between ISI mitigation and noise amplification, implying the existence of an optimal derivative order that minimizes BER. The derivative order optimization problem is addressed for fixed threshold detectors, through the derivations of the theoretical BER expressions of $\boldsymbol{D}^{m}$-FSTD and $\boldsymbol{D}^{m}$-MaTD pairs. Furthermore, an SINR-like objective function is proposed to optimize $m$ for $\boldsymbol{D}^{m}$-FSTD. Numerical results confirm the accuracy of the derived expressions, demonstrate the efficacy of the $\boldsymbol{D}^{m}$ operator in ISI mitigation, and show that said ISI mitigation decreases the needed memory window of memory-aided detectors. Overall, the proposed $\boldsymbol{D}^m$ operator is shown to be a computationally very cheap strategy that provides a powerful ISI mitigation. Given the transmitter is able to handle large transmission powers to alleviate the effects of noise amplification, the derivative operator allows for achieving considerably higher data rates, while still preserving a reliable communication link.

%\balance

\appendices
\section{Clark's Approximation for MaTD}
\label{ap:clark} 

\par Let $X_i = y_{L,(m)}[i]$ where $\boldsymbol{y}_{L,(m)} = \boldsymbol{D}^m \boldsymbol{y}_{L}$. Then, $X_1,\dots,X_{N-m}$ are correlated and differently distributed random variables whose maximum's distribution is of interest. In its first iteration, Clark's approximation finds the mean and the variance of $\max(X_1,X_2)$. Afterwards, $\max(X_1,X_2)$ is approximated as a Gaussian with said mean and variance\footnote{Note that maximum of two Gaussians is not a Gaussian itself. However, Clark's approximation considers it as such in order to devise its recursive strategy \cite{ClarkApprox,Clark_notgaus}.}. Then, in the second iteration, $\max(X_1,X_2,X_3) = \max[\max(X_1,X_2),X_3]$ simply becomes the maximum of two ``Gaussians", which is handled in a similar way to the first iteration. The process keeps iterating until the $n^{th}$ random variable. 

\par Herein, we present the method of finding the mean and the variance of $\max(X_1,X_2)$. Let $X_1 \sim \mathcal{N}(\mu_1,\sigma^2_1)$ and $X_2 \sim \mathcal{N}(\mu_2,\sigma^2_2)$. Note that by definition, $\mu_i$ is the $i^{th}$ entry of $\boldsymbol{\mu}_{L,(m)}$, and $\sigma^2_i$ is the $i^{th}$ diagonal entry of $\boldsymbol{\Sigma}_{L,(m)}$. We define two auxiliary parameters $a$ and $\alpha$ to be
\begin{equation}
\label{eq:alphas}
    \begin{split}
        a &= \sqrt{\sigma^2_{1} + \sigma_{2}^2 - 2\sigma_{1,2}} \\
        \alpha &= \frac{\mu_1 - \mu_2}{a}.
    \end{split}
\end{equation}
Let $\nu_1$ and $\nu_2$ denote the first and second moments of $\max(X_1,X_2)$. Then, 
\begin{equation}
\label{eq:clark_moments}
\begin{split}
    \nu_1 &= \mu_1 Q(-\alpha) + \mu_2 Q(\alpha) + a \psi(\alpha) \\
    \nu_2 &= (\mu^2_1 + \sigma^2_{1}) Q(-\alpha) + (\mu^2_2 + \sigma^2_{2}) Q(\alpha) + (\mu_1 + \mu_2) a \psi(\alpha),
\end{split}
\end{equation}
where $\psi(\cdot)$ denotes the standard normal PDF. Lastly, using $\nu_1$ and $\nu_2$, we approximate $\max(X_1,X_2) \sim \mathcal{N}(\nu_1,\nu_2-\nu^2_1)$, completing the iteration.

\bibliographystyle{IEEEtran}
\bibliography{refs_derivative}

\end{document}